\documentclass[floats,floatfix,showpacs,preprintnumbers,amssymb,prd,superscriptaddress,nofootinbib,nolongbibliography,reprint]{revtex4-1}
\usepackage[dvipdfmx]{graphicx}

\usepackage{subfigure}
\usepackage{bm}
\usepackage{mathrsfs}
\usepackage{amsmath}
\usepackage{cases}

\def\be{\begin{equation}}
\def\ee{\end{equation}}
\newcommand{\beq}{\begin{eqnarray}}
\newcommand{\eeq}{\end{eqnarray}} 
\usepackage{color}
\usepackage{ulem}
\usepackage[linktocpage=true]{hyperref}

\begin{document}
\title{Tidal response beyond vacuum General Relativity with a canonical definition}

\author{Takuya Katagiri}
\affiliation{Center of Gravity, Niels Bohr Institute, Blegdamsvej 17, 2100 Copenhagen, Denmark}

\author{Vitor Cardoso}
\affiliation{Center of Gravity, Niels Bohr Institute, Blegdamsvej 17, 2100 Copenhagen, Denmark}

\affiliation{CENTRA, Departamento de F\'{\i}sica, Instituto Superior T\'ecnico -- IST, Universidade de Lisboa -- UL,
Avenida Rovisco Pais 1, 1049-001 Lisboa, Portugal}

\author{Tact Ikeda}
\affiliation{Department of Physics, Rikkyo University, Toshima, Tokyo 171-8501, Japan}

\author{Kent Yagi}
\affiliation{Department of Physics, University of Virginia, Charlottesville, Virginia 22904, USA}

\date{\today}

\begin{abstract}
Tidal effects on compact objects provide profound theoretical insights into the structure of the field equations, and are wonderful probes of the equation of state of matter, of the nature of black holes and of the underlying theory of gravity. The natural framework for understanding tides is a perturbative scheme. Here, we point out ambiguities in determining tidal response functions within such a framework, which may lead to bias in constraining physical parameters with gravitational-wave observations if the computed quantities are not properly linked to observables. We propose a Canonical Tidal Response Function (CTRF) definition to compare values of tides in theories beyond vacuum General Relativity in a unified manner. As an example, we provide black hole tidal response functions, including both conservative and dissipative pieces, in various theories of gravity. Tidal dissipation Love numbers for black holes are derived here for the first time in most of the non-Einsteinian theories considered in this paper.
\end{abstract}

\preprint{RUP-24-19}
\maketitle

\section{Introduction}

\subsection{Tidal effects and fundamental physics}

Tides are promising and powerful probes to understand the nature of compact objects, the underlying theory of gravity, and astrophysical environments around binary systems. Tides between binary constituents affect its binary dynamics, leaving detectable imprints in gravitational waveforms. To connect theoretical predictions on relativistic tidal effects with future observational outcomes, their accurate and unambiguous modeling is essential.

Tidal effects are encoded in a set of {\it tidal response functions}. Their leading conservative part is quantified by {\it tidal Love numbers}~(TLNs)~\cite{poisson_will_2014}, whose leading term appears at $5$th post-Newtonian (PN) order in a gravitational waveform for a compact binary inspiral~\cite{Flanagan:2007ix,Flanagan:2007ix,Vines:2010ca,Abac:2023ujg,Narikawa:2023deu}. 
Within General Relativity (GR), black holes (BHs) have vanishing TLNs~\cite{Binnington:2009bb,Landry:2015zfa,LeTiec:2020spy,LeTiec:2020bos,Charalambous:2021mea,Chia:2020yla,Poisson:2020vap,Poisson:2021yau}, a property that seems to be specific to vacuum GR: BHs in alternative theories of gravity or non-vacuum environments generally have non-vanishing TLNs~\cite{Cardoso:2017cfl,Cardoso:2018ptl,Kol:2011vg,Cardoso:2019vof,DeLuca:2022tkm,vanGemeren:2023rhh,Katagiri:2023umb}. 
The precise measurement of the phase of gravitational waves from compact binaries allows one to infer their tidal properties, and therefore to constrain deviations from GR~\cite{Cardoso:2017cfl,Maselli:2018fay},
to probe the environment where BHs evolve~\cite{Cardoso:2019upw,Cardoso:2021wlq,Katagiri:2023yzm,Cannizzaro:2024fpz,DeLuca:2024uju,Arana:2024kaz} or even to place constraints on fundamental degrees of freedom~\cite{DeLuca:2021ite,DeLuca:2022xlz}. TLNs thus can inform us on a variety of foundational issues. All observations to date are consistent with vacuum GR~\cite{Narikawa:2021pak,Chia:2023tle}. Next generation interferometers will be able to put stringent constraints on BH TLNs~\cite{Maselli:2018fay,Pani:2019cyc,Puecher:2023twf,DeLuca:2021ite,DeLuca:2022xlz}. Driven by their potential to convey information about gravity in the strong and dynamical field regime, there have been a number of studies on conservative effects up to the next-to-leading order of a low-frequency expansion or nonlinear order of tidal perturbations towards highly accurate and precise modeling of the dynamical or nonlinear tidal deformation in late inspiralling stages~\cite{Hinderer:2016eia,Steinhoff:2016rfi,Pratten:2019sed,Poisson:2020vap,Pratten:2021pro,Pitre:2023xsr,HegadeKR:2024agt,Katagiri:2024wbg}.

In neutron star physics, TLNs depend on the nuclear matter equation of state and the underlying gravitational theory~\cite{Hinderer:2007mb,Damour:2009vw,Binnington:2009bb}. The first detection of gravitational waves from a binary neutron star merger, GW170817, has constrained the tidal deformabilities, leading to measurements of radii and constraining nuclear matter equations of state at supernuclear densities~\cite{LIGOScientific:2017vwq,LIGOScientific:2018cki,LIGOScientific:2018hze}. The discovery of universal (``I-Love-Q'') relations among the moment of inertia, the TLNs, and the quadrupole moment  reveals intriguing and profound theoretical structures underlying tidal physics for neutron stars, and provides a useful tool to test GR in the strong-field regime~\cite{Yagi:2013bca,Yagi:2013awa,Yagi:2016bkt,Silva:2020acr}. Similar relations that connect TLNs at different multipole orders have also been found~\cite{Yagi:2013sva}, breaking the degeneracy between tidal parameters in data analyses. In the context of cosmology, the relations found in Refs.~\cite{Yagi:2015pkc,Yagi:2016qmr}, coined binary Love relations, open a new avenue to measure the Hubble constant without an electromagnetic counterpart in an equation-of-state insensitive way~\cite{Chatterjee:2021xrm}. On the other hand, if the cosmology is known, binary Love relations can be used to break the distance-inclination degeneracy~\cite{Xie:2022brn}.

Aside from the observational relevance, theoretical studies on BH TLNs unveil a rich structure of hidden symmetries behind the vanishing of TLNs in vacuum, four-dimensional GR~\cite{Porto:2016zng,Penna:2018gfx,Hui:2021vcv,Charalambous:2021kcz,Katagiri:2022vyz,Charalambous:2022rre,BenAchour:2022uqo,Hui:2022vbh}. Follow-up analyses demonstrate their emergence in a wider class of spherically symmetric BHs~\cite{Charalambous:2024tdj,Sharma:2024hlz,Rai:2024lho}. Higher-dimensional Schwarzschild BHs exhibit nonzero TLNs and display scale-dependent logarithmic behaviors in their tidal perturbations~\cite{Kol:2011vg,Cardoso:2019vof,Hui:2020xxx,Pereniguez:2021xcj,Charalambous:2024tdj}. Within the AdS/CFT correspondence, TLNs of asymptotically anti-de Sitter BHs are related to polarization coefficients of plasma~\cite{Emparan:2017qxd}.

\begin{table}[ht!]
       \begin{tabular}{|c|c|c|} \hline
    BH TLNs & $\ell=2$ & $\ell=3$  \\  \hline
   GR & $0,0$ & $0,0 $ \\ \hline
   E-M & ${\cal O}\left(q^3 \right), {\cal O}\left(q^3 \right)$ & ${\cal O}\left(q^3 \right),{\cal O}\left(q^3 \right)$\\  \hline
   dCS & $0, 0.034\zeta_{\rm dCS}^2$ & $0, 0.10\zeta_{\rm dCS}^2$ \\  \hline
   EdGB & $-0.75\zeta_{\rm dGB}^2, \frac{361}{480}\zeta_{\rm dGB}^2$ & $-0.043\zeta_{\rm dGB}^2, \frac{188483}{376320}\zeta_{\rm dGB}^2$ \\  \hline
   Braneworld & $\frac{1}{32}\zeta_{\rm B}, \frac{1}{216}\zeta_{\rm B}$ & $-\frac{83}{4608}\zeta_{\rm B}, \frac{1}{2304}\zeta_{\rm B}$  \\  \hline
   EFT~($\epsilon_1$) & $-\frac{63}{50}\epsilon_1, \frac{27}{50}\epsilon_1 $ & $-\frac{99}{8}\epsilon_1, \frac{417}{56}\epsilon_1 $ \\  \hline
  EFT~($\epsilon_2$) & ${\cal O} (\epsilon_2^2), -\frac{12}{5}\epsilon_2$ & ${\cal O} (\epsilon_2^2), -\frac{129}{7}\epsilon_2$  \\  \hline
 \end{tabular}
\caption{$\ell$th-multipole order TLNs for non-spinning BHs, for even and odd parity. The theories above are precisely given in Appendix~\ref{Appendix:model}. Note that $q, \zeta_{\rm dCS}, \zeta_{\rm dGB}, \zeta_{\rm B}, \epsilon_1, \epsilon_2$ are dimensionless coupling constants in the theories. The above values are consistent with the results for GR~\cite{Binnington:2009bb,LeTiec:2020spy,LeTiec:2020bos,Charalambous:2021mea,Chia:2020yla,Poisson:2020vap,Poisson:2021yau,Katagiri:2024wbg}, Einstein-Maxwell~\cite{Cardoso:2017cfl,Poisson:2021yau}, dynamical Chern-Simons~(dCS)~\cite{Cardoso:2017cfl}, braneworld~\cite{Tan:2020hog}, and effective field theory~(EFT) framework~\cite{Cardoso:2018ptl} together with the parametrized formalism~\cite{Katagiri:2023umb}, while not with findings for braneworld in another analysis~\cite{Chakravarti:2018vlt}. We explain the cause for this tension in Sec.~\ref{Sec:Application}. Note the difference in the conventions with Refs.~\cite{Cardoso:2017cfl,Tan:2020hog,Cardoso:2018ptl}~(see footnote~\ref{footnote:conventions}). The results for the odd sector of dCS and the even sector of Einstein-dilaton-Gauss-Bonnet~(EdGB) are approximate estimations~(see Appendix~\ref{Appendix:model}). To our knowledge, the BH TLNs in EdGB gravity are computed for the first time.}
\label{table:TLNs}
\end{table}
\begin{table*}
    \centering
\begin{tabular}{|c|c|c|} \hline
    Deviation of BH TDNs& $\ell=2$ & $\ell=3$ \\  \hline
   E-M & ${\cal O}\left(q^3\right), {\cal O}\left(q^3\right)$ & $-\frac{1}{2400}q^2 ,-\frac{1}{2400}q^2$  \\  \hline
   dCS & $0, 0.16\zeta_{\rm dCS}^2 $ & $0 ,0.68\zeta_{\rm dCS}^2$  \\  \hline
   EdGB & $-5.57\zeta_{\rm dGB}^2,\frac{45331}{14400}\zeta_{\rm dGB}^2$ & $-0.33\zeta_{\rm dGB}^2,\left(\frac{145027853}{3763200}-\frac{25\pi^2}{7}\right)\zeta_{\rm dGB}^2 $  \\  \hline
   Braneworld & $-0.014\zeta_{\rm B}, -\frac{17}{1296}\zeta_{\rm B}$ & $-0.049\zeta_{\rm B},\frac{23}{23040}\zeta_{\rm B}$  \\  \hline
   EFT~($\epsilon_1$) & $-\frac{4121}{2000}\epsilon_1,\frac{97919}{42000}\epsilon_1$ & $-\frac{31807}{960} \epsilon_1, \frac{2300293}{47040}\epsilon_1$ \\  \hline
  EFT~($\epsilon_2$)   & ${\cal O}\left(\epsilon_2^2\right), -\frac{2606}{175}\epsilon_2$ & ${\cal O}\left(\epsilon_2^2\right),-\frac{59557}{490}\epsilon_2$ \\  \hline
 \end{tabular}
\caption{The deviation of $\ell$th-multipole order TDNs, $\nu_\ell^\pm$, for non-spinning BHs from GR values,~$\nu_\ell^{\pm (0)}$ given by Eq.~\eqref{eq:GRTDNs}: $\nu_\ell^{\pm}-\nu_\ell^{\pm (0)}$. The results for the odd sector of dCS and the even sectors of EdGB and braneworld are approximate estimations~(see Appendix~\ref{Appendix:model}). To our knowledge, this is the first computation for TDNs in non-GR theories.\footnote{Tidal heating of the braneworld BH was investigated in Refs.~\cite{Chakraborty:2021gdf,Krishnendu:2024jkj} but these works ignored coupling between the gravitational sector and the effective matter sector arising from the bulk Weyl tensor at a perturbative level. Moreover, they did not compute TDNs.}}
    \label{table:TDNs}
\end{table*}
The dissipative process caused by tides is referred to as tidal heating~\cite{Poisson:2004cw,Cardoso:2012zn,Chatziioannou:2012gq,poisson_will_2014}, often called tidal dissipation~\cite{Chia:2020yla,Charalambous:2021mea}, whereby energy and angular momentum  are transferred between the object and the tidal environment during the binary evolution. 
The leading dissipative tidal effects are measured by a set of {\it tidal dissipation numbers}~(TDNs). This effect first appears at $2.5$ ($4$) PN order in the gravitational waveform for rotating (non-rotating) objects~\cite{Poisson:1994yf,Tagoshi:1997jy,Alvi:2001mx}. For BHs, dissipation is sourced by absorption at the horizon. The presence of dissipative mechanisms could serve as a discriminator for hypothetical horizonless compact objects with Planckian corrections at the horizon scale~\cite{Maselli:2017cmm,Cardoso:2019nis}. Precise tests of the horizon presence or absence may be possible in third-generation detectors~\cite{Mukherjee:2022wws}. The value of TDNs depends on the underlying theories of gravity, allowing us to test GR~\cite{Chia:2024bwc}. Tidal heating is particularly significant in extreme mass ratio inspirals, providing unprecedented opportunities to test strong-field gravity with future detectors~\cite{Hughes:2001jr,Price:2001un,Martel:2003jj,Cardoso:2019nis,Datta:2019euh,Datta:2019epe,Datta:2020rvo,Datta:2024vll}. Quantifying the tidal-heating effects allows us to infer the nature of lower mass-gap binaries~\cite{Datta:2020gem}. Internal mechanisms such as shear viscosity of ultra-dense nuclear matter can cause dissipation in neutron stars~\cite{Ripley:2023lsq,HegadeKR:2024slr}. Precise measurements of TDNs thus lead to constraints on the properties of nuclear matter in extreme environments. 

\subsection{Ambiguity in concept of relativistic external tidal field}
Tidal response functions are defined from the linear relation connecting the external tidal field to the induced response. In spite of this simple statement, the operational relativistic definition is subtle, due to ambiguities in the determination of those values. 

In Newtonian gravity, the external tidal field and response field are unambiguously defined, and hence, one can specify a tidal response function uniquely~\cite{poisson_will_2014}. In a relativistic framework, on the other hand, the concept of an external field is ambiguous because of mixing of the body's own contribution with an external source in a tidally perturbed metric. This allows arbitrary shifts of a tidal response function with redefinition of an ``external tidal field"~\cite{Fang:2005qq,Gralla:2017djj,Poisson:2020vap,Katagiri:2024wbg}. In a vacuum GR setup, a definition based on the analytic properties of hypergeometric functions is commonly used, leading to vanishing of TLNs of Kerr BHs~\cite{Binnington:2009bb,Hui:2020xxx,LeTiec:2020spy,LeTiec:2020bos,Charalambous:2021mea,Chia:2020yla}.

\subsection{This work: modeling a relativistic tidal response under a unified definition}
In this work, we point out that the absence of a ``canonical'' definition of tidal response functions leads to an important ambiguity in the determination of such functions in theories that extend vacuum GR. The ambiguity arises from the degrees of freedom in the functional form of particular solutions of the inhomogeneous equation that appears in the perturbative expansion scheme, in terms of the small parameter controlling the deviation from vacuum GR.\footnote{Essentially, the same issue is investigated in the context of dynamical tides~\cite{HegadeKR:2024agt,Katagiri:2024wbg}.} This degree of freedom translates into the freedom to decompose the integration constant, determined by an inner boundary condition, into external tidal and  induced response pieces. To be clear, a tidally deformed metric is free from the ambiguity. However, different values of a tidal response function based on implicit assumptions in the definition may lead to a bias in constraining TLNs and TDNs from measured tidal deformabilities with gravitational-wave observations if the computed quantity is not properly linked to observables. Details can be found in Sec.~\ref{Sec:TheoryAndAmbiguity} and in Ref.~\cite{Katagiri:2024wbg}.

We therefore propose a systematic procedure to fix a tidal response function in Sec.~\ref{Sec:Definition}, allowing one to compute TLNs and TDNs of compact objects~(not necessarily BHs) in non-GR and/or non-vacuum setups as differences from the values of GR BHs in a unified manner. As an example, we provide BH values in various alternative theories of gravity within the unified definition, which are summarized in Tables~\ref{table:TLNs} and~\ref{table:TDNs}, updating the previous work focused on calculating TLNs for non-GR BHs and exotic compact objects~\cite{Cardoso:2017cfl}.\footnote{\label{footnote:conventions} It should be noted that the literature~\cite{Cardoso:2017cfl,Tan:2020hog,Cardoso:2018ptl} adopt the different convention by Cardoso, Franzin, Maselli, Pani, and Raposo~(CFMPR) from ours:
\begin{align}
    \kappa_{\ell,{\rm CFMPR}}^\pm=\left(\frac{r_0}{M}\right)^{2\ell+1} \kappa_\ell^\pm,
\end{align}
where $M$ and $r_0$ are mass and radius of the object.}   We discuss discrepancies of our results with the literature~\cite{Chakravarti:2018vlt} in Sec.~\ref{Sec:Application}. We conclude this work in Sec.~\ref{Sec:Conclusion}. We adopt geometrical units~$c=G=1$ throughout the paper.

\section{Theory of tidal response and ambiguities}\label{Sec:TheoryAndAmbiguity}
 
\subsection{Relativistic theory of tidal response}
We now briefly review the relativistic theory of tidal response based on Refs.~\cite{Taylor:2008xy,Poisson:2009qj,poisson_will_2014,Poisson:2018qqd,Poisson:2020vap,Katagiri:2024wbg}. Consider an inspiralling binary system with a large separation as compared to the radii of the constituents. We describe tidal effects on one object, caused by the other, within a linear response theory.

The description of a tidal response relies on matched asymptotic expansions of a body metric that describes the neighborhood of the body in vacuum with a PN metric that describes an external universe except for the vicinity of the body~\cite{Taylor:2008xy,Poisson:2018qqd}. 
The large separation condition,~$r_{b}\gg M$, where $r_{b}$ and $M$ denote the orbital separation and the body mass, respectively, is translated into the condition, $M\ll {\cal R}$, for the characteristic length scale of the external universe,~${\cal R}$.\footnote{Assuming that the body is moving in a quasi-circular orbit around the other, ${\cal R}$ is of order of the wavelength of the gravitational wave being emitted from the binary. Then, ${\cal R}/M\sim (r_b/M_{\rm tot})^{3/2}$ holds for the total mass~$M_{\rm tot}$ in the system. When $r_b\gg M$, then ${\cal R}\gg M$.}  This allows one to separate the outer region of the body into two regions: the first one is a {\it body zone}~$(M \lesssim r\ll {\cal R})$, which is a neighborhood of the body with a fully GR description; the other is a {\it post-Newtonian~(PN) zone}~($M\ll r)$, where gravity is weak and is described by a PN expansion of the global metric. 
The body zone extends up to the PN zone; the PN zone extends up to the body zone. There is an overlapping region, $M\ll r\ll {\cal R}$, between the two regions. We call this region a {\it buffer zone}. Asymptotically matching the body metric and the PN metric in the buffer zone in the same gauge and the same coordinates allows one to construct the whole spacetime metric approximately, thereby describing the tidal response of the body. 

The body (exterior) metric is a solution of the linearized Einstein equations. The large-distance asymptotic expansion of the metric in the buffer zone has information on the multipolar structure of the body, called {\it induced multipole moments}~\cite{Hansen:1974zz,Geroch:1970cd,RevModPhys.52.299,1983GReGr..15..737G,Poisson:2020vap}. Before matching, the body metric does not have information on an external source. Therefore, the body metric contains unspecified coefficients. 

In the PN zone, the body can be seen as a point mass with an unspecified multipolar structure, moving in an ambient weak field. The PN metric has information on tidal environments created by an external source but does not have the detailed description, induced multipole moments, of the body before matching.  
Assuming that the point mass follows a timelike geodesic~$\gamma$, the information on tidal environments is encoded into {\it tidal moments} as follows~\cite{Zhang:1986cpa,Taylor:2008xy,Poisson:2009qj}: first, the metric of the external universe is expanded in the local inertial frame constructed from an orthogonal basis that is parallel-transported along $\gamma$. Then, the components of the Weyl tensor and its covariant derivatives on $\gamma$ in the local frame define tidal moments:
\begin{align}
    {\cal E}_L\left(t\right):=&\frac{1}{\left(\ell-2\right)!}\left[C_{0i_10i_2|i_3\cdots i_\ell}\right]^{\rm STF},\label{eq:eletidal}\\
     {\cal B}_L\left(t\right):=&\frac{3}{2\left(\ell+1\right)\left(\ell-2\right)!}\left[\varepsilon_{i_1 jk}C^{jk}_{~~i_20|i_3\cdots i_\ell}\right]^{\rm STF},\label{eq:magtidal}
\end{align}
where $t$ is a proper time along $\gamma$; $\varepsilon$ is the permutation symbol; $\ell$ denotes $\ell$th order in the multipole expansion; $|$ is a covariant derivative projected onto the local inertial frame; $L$ stands for an $L$~collection of indices~$i$; STF means symmetric and tracefree with any pair. 
Here, $ {\cal E}_L$ and $ {\cal B}_L$ are called {\it electric-type tidal moments} and {\it magnetic-type tidal moments}, respectively. 

With the matching of a body metric and a PN metric, the PN information determines the tidal moments in the body zone, and the body-zone information determines the multipole moments in the PN zone~\cite{Taylor:2008xy,Poisson:2009qj,Poisson:2018qqd,Poisson:2020vap,Pitre:2023xsr,HegadeKR:2024agt}. Within linear response theory, electric-type induced multipole moments~$I_{L}^+$ are linearly related with electric-type tidal moments~${\cal E}_{L}$ as
%
%
\begin{align}
    I_{L}^+\left(t\right)=-\frac{2}{\left(2\ell-1\right)!!}r_0^{2\ell+1} \int_{-\infty}^\infty {\cal F}_\ell^+\left(t-t'\right) {\cal E}_{L}\left(t'\right) dt',
\end{align}
where $r_0$ is the body radius.  Here, ${\cal F}_\ell^+$ is called an {\it electric-type tidal response function}. The same relation holds for the magnetic-type piece, defining a {\it magnetic-type tidal response function}~${\cal F}_\ell^-$. The tidal response function depends on the internal structure of the body as well as the underlying theory of gravity.

In the Fourier domain, assuming the low frequency~$\omega M \ll 1$, the tidal response function is expanded as
\begin{align}
    {\cal F}^\pm_{\ell}\left(\omega\right)=\kappa_\ell^\pm +i \omega M \nu_\ell^\pm+{\cal O}((\omega M)^2).
\end{align}
 Here, the leading coefficients in the conservative and dissipative parts are called {\it electric-type/magnetic-type TLNs} and {\it electric-type/magnetic-type TDNs}, respectively.

\subsection{Ambiguities in determining tidal responses}

\subsubsection{Unambiguous tides in Newtonian gravity}
Tidal environments around a body are characterized by tidal moments, independent of the details of the body. In Newtonian gravity, tidal moments are defined from the Taylor expansion of the gravitational potential of an external source responsible for the tidal environment around the body. Induced multipole moments are defined from an integration involving matter density over the volume of spatial domain occupied by the body. The external tidal field and an induced response field are defined unambiguously. An $\ell$th polar external tidal field and induced response field can be extracted from the terms in the gravitational potential that scale as $r^\ell$ and $r^{-\ell-1}$~($r=(x_i x^i)^{1/2}$ in Cartesian coordinates~$x^i$), respectively, derived from the Laplace equation~\cite{poisson_will_2014}. 

\subsubsection{Ambiguous external tides in relativistic theories}
The unambiguous description in Newtonian gravity does not carry over to the relativistic framework. There are ambiguities in the concept of an external tidal field, for the following reasons~\cite{Taylor:2008xy,Poisson:2020vap,Katagiri:2024wbg}. First, a body has, in fact, a finite size, and hence, the worldline is replaced by a worldtube. Second, a body does not follow a spacetime geodesic in general. Third, the Weyl tensor in the buffer zone has the body's own contribution as well, not only that of the external source. Therefore, the definition of tidal moments through Eqs.~\eqref{eq:eletidal}--~\eqref{eq:magtidal} has a robust interpretation only in the masssless limit of the body metric, and the concept of an external field is ambiguous. There is no unambiguous way to decompose the body metric into external tidal and induced response pieces. We show this with a clear, full non-linear example of a BH binary, described by the Majumdar-Papapetrou metric in Appendix~\ref{sec:MP}.

We further express the ambiguity between a tidal field and a response field as follows. The perturbation to the effective gravitational potential read off from the asymptotic expansion of the body metric in the buffer zone, in the harmonic radial coordinate~$\bar{r}$, takes the form of
\begin{align}
    \delta U_{\rm eff}\sim&  \big[{\cal E}_{L} \bar{r}^\ell \left\{1+{\cal O}\left(M/\bar{r}\right)\right\}\label{eq:deltaeffU}\\
    &+\frac{I_{L}^+}{\bar{r}^{\ell+1}}\left\{1+{\cal O}\left(M/\bar{r}\right)\right\}\big]\Omega^L e^{-i \omega t},\nonumber
\end{align}
where $\Omega^a=[\sin \theta \cos \varphi,\sin\theta\sin\varphi,\cos\theta]$. The body's finite-size effect introduces subleading corrections of the order of $M/\bar{r}$ to the $\bar{r}^\ell$ and $\bar{r}^{-\ell-1}$ terms. The subleading corrections to the Newtonian tidal field vanish when $M\to 0$, and then, ${\cal E}_{L}\bar{r}^\ell$ acquires the meaning of an external tidal field; however, the series itself in general contains the body's own information, not only that of the tidal environment. Equation~\eqref{eq:deltaeffU} implies that the correction in the first line may have a term proportional to $\bar{r}^{-2\ell-1}$, giving rise to the term~${\cal E}_{L}\delta_L/\bar{r}^{\ell+1}$, with $\delta_L$ an ${\cal O}(1)$ constant, which is degenerate with the term of $I_{L}^+/\bar{r}^{\ell+1}$ in the second line. One can then redefine $I_{L}^+$ into $\tilde{I}_L^+=I_{L}^++{\cal E}_{L}\delta_L$ by taking the shift into account. The resulting tidal response function is altered due to this redefinition~\cite{Fang:2005qq,Gralla:2017djj,Poisson:2020vap}.

In vacuum GR, an analytic continuation of $\ell$ has been proposed as a means to extract a tidal response function without such subtlety~\cite{Kol:2011vg,Hui:2020xxx,Creci:2021rkz,Charalambous:2021mea}. In fact, the prescription makes distinct growing and decaying modes in the radial coordinate. This prescription corresponds to establishing a ``standard" definition within vacuum GR, based on the analytic properties of hypergeometric functions. Using such functions as basis functions, one can show that Kerr BH TLNs vanish~\cite{Hui:2020xxx,Chia:2020yla,LeTiec:2020bos,LeTiec:2020spy,Katagiri:2024wbg}. However, if one uses another basis that leads to the shift of the coefficient of the $\bar{r}^{-\ell-1}$ term, then the resulting values change~\cite{Gralla:2017djj,Poisson:2020vap,Katagiri:2024wbg}. We note that, in a perturbative framework of non-vacuum GR setups, the analytic continuation, such as the prescription in Ref.~\cite{Barura:2024uog}, is no longer useful as the assignment of integration constants of particular solutions to tidal and response pieces requires a subjective definition of a tidal field.

\subsection{Observables are ambiguity-free}
Although the tidal response function depends on the choice of basis functions in the decomposition of tidal and response fields, the tidally deformed metric is free from the ambiguity, given an inner boundary condition. This can be understood from the fact that imposing an inner boundary condition determines one of the two integration constants in the general solution, and hence, the functional form of the metric components is uniquely specified. The aforementioned ambiguity corresponds to degrees of freedom to decompose the integration constant determined by an inner boundary condition into external tidal and induced response pieces.

The description of any dynamics on the tidally deformed background is thus unambiguous. Therefore, any choices of the definition of a tidal response function have no impact on the prediction on observable quantities, i.e., gravitational waves. One can set the free parameter arbitrarily, thereby obtaining different values for the tidal response function; once a common convention is adopted, all results can be meaningfully compared even in the absence of a matching scheme between the tidal response function computed in the body zone and observables.

\section{Definition of a canonical tidal response function}\label{Sec:Definition}
We first explain a useful perturbative expansion scheme when considering an extension of vacuum GR. We then propose a canonical tidal response function (CTRF) that allows us to determine tidal responses for any compact objects in a unified and simple manner.

\subsection{Perturbative expansion}
Consider now an extension of vacuum GR, parametrized by some coupling constant $\epsilon$. The coupling constant is assumed to be small compared to any other scales in the binary. 
We also assume low frequencies compared to~$M$, allowing two-parameter expansions of perturbations in $(\epsilon, \omega M)$. Here, we focus on the even-parity and odd-parity sectors of metric perturbations in the Regge-Wheeler gauge~\cite{PhysRev.108.1063}. Appendix~\ref{App:GRsolution} provides the low-frequency expansions of metric, vector-field, and scalar-field perturbations on a Schwarzschild background in vacuum GR, corresponding to the GR limit of the analysis in this section. The material for the computation of tidal perturbations in particular models is provided in Appendix~\ref{Appendix:model}.

The even-parity and odd-parity metric perturbations are fully determined once, in $(t,r,\theta,\varphi)$ coordinates,  one specifies the $tt$ and $t\varphi$ components, respectively. With spherical harmonic decomposition in the Fourier domain, the radial parts of the $tt$ and $t\varphi$ components are determined by the following expansions of functions of $r$ in $\epsilon$ and $\omega M$:
\begin{align}
H \simeq &H^{(0,0)} +\epsilon H^{(1,0)} +\epsilon^2 H^{(2,0)} \nonumber\\
&+\omega M \left( H^{(0,1)} +\epsilon  H^{(1,1)}+\epsilon^2  H^{(2,1)} \right),\\
h \simeq &h^{(0,0)} +\epsilon h^{(1,0)} +\epsilon^2 h^{(2,0)} \nonumber\\
&+\omega M \left( h^{(0,1)} +\epsilon  h^{(1,1)}+\epsilon^2  h^{(2,1)} \right).
\end{align}
Here, the superscripts~$(i,j)$ count $i$th and $j$th order of $\epsilon$ and $\omega M$, respectively. 

The expansion of the perturbation equations for $H$ and $h$ take the forms,
\begin{align}
    {\cal L}_H \left[H^{(0,i)}\right]=&0,~{\cal L}_H \left[H^{(1,i)}\right]={\cal S}_H^{(1,i)},~{\cal L}_H \left[H^{(2,i)}\right]={\cal S}_H^{(2,i)},\label{eq:H01i}\\
        {\cal L}_h \left[h^{(0,i)}\right]=&0,~{\cal L}_h \left[h^{(1,i)}\right]={\cal S}_h^{(1,i)},~{\cal L}_h \left[h^{(2,i)}\right]={\cal S}_h^{(2,i)},\label{eq:h01i}
\end{align}
where ${\cal L}_{H/h}$ are defined in Eqs.~\eqref{eq:LHGR} and~\eqref{eq:LhGR}, respectively. The source terms~${\cal S}_{H/h}^{(i,j)}$ arise from non-vacuum and/or non-GR effects and are model dependent. Equations~\eqref{eq:H01i} and~\eqref{eq:h01i} imply that the general solutions at ${\cal O}(\omega M)$ take the same form as static perturbations. This can be understood from the fact that perturbation equations on a spherically symmetric background do not contain ${\cal O}(\omega^1)$ terms.

The general solutions at $(0,i)$th order take the form,
\begin{align}
H^{(0,i)}=\mathbb{E}_H^{(0,i)}  H_T +\mathbb{I}_H^{(0,i)} H_R,~h^{(0,i)}  =\mathbb{E}_h^{(0,i)}  h_T +\mathbb{I}_h^{(0,i)}  h_R, \label{eq:HhGR}
\end{align}
where $\mathbb{E}_{H/h}^{(0,i)}$ and $\mathbb{I}_{H/h}^{(0,i)} $ are integration constants; $H_{T/R}$ and $h_{T/R}$ are defined in Eqs.~\eqref{eq:HT}-\eqref{eq:hR}. The subscripts~$T$ and $R$ indicate that their leading terms in the buffer zone will be matched with Newtonian tidal and induced response pieces, respectively, in matching of the perturbed metric with a PN metric~\cite{Katagiri:2024wbg}.

The general solutions at $(1,i)$th and $(2,i)$th order consist of the homogeneous piece~\eqref{eq:HhGR} and particular solutions:
\begin{align}
       H^{(1,i)}= &\mathbb{E}_H^{(1,i)}H_T+\mathbb{I}_H^{(1,i)}H_R+P_H^{(1,i)},\label{eq:H1i}\\
       h^{(1,i)}= &\mathbb{E}_h^{(1,i)}h_T+\mathbb{I}_h^{(1,i)}h_R+P_h^{(1,i)},\label{eq:h1i}
\end{align}
and
\begin{align}
       H^{(2,i)}= &\mathbb{E}_H^{(2,i)}H_T+\mathbb{I}_H^{(2,i)}H_R+P_H^{(2,i)},\label{eq:H2i}\\
       h^{(2,i)}= &\mathbb{E}_h^{(2,i)}h_T+\mathbb{I}_h^{(2,i)}h_R+P_h^{(2,i)},\label{eq:h2i}
\end{align}
where $P_{H/h}^{(i,j)}=P_{H/h}^{(i,j)}(r)$ are particular solutions; $\mathbb{E}_{H/h}^{(i,j)}$ and $\mathbb{I}_{H/h}^{(i,j)}$ are integration constants. We emphasize that the functional form of particular solutions is not uniquely specified, allowing arbitrary shifts of the coefficients in front of the homogeneous solutions. This freedom directly leads to ambiguities of tidal response functions in a perturbative framework~\cite{HegadeKR:2024agt,Katagiri:2024wbg}.

\subsection{Quantication of tidal responses}
\subsubsection{Bare tidal response function}
In vacuum GR, matching a body metric with a PN metric relates $\mathbb{E}_{H/h}^{(0,j)}$ and $\mathbb{I}_{H/h}^{(0,j)}$ to tidal moments and tidally induced multipole moments, respectively~\cite{Poisson:2020vap,Pitre:2023xsr,HegadeKR:2024agt}. We assume that the same identification works even in the current framework, and then, regard $\mathbb{E}_{H/h}^{(i,j)}$ and $\mathbb{I}_{H/h}^{(i,j)}$ with $i=1,2$ as corrections to the tidal moments and the multipole moments, respectively. The details of the matching procedure are provided in Refs.~\cite{Poisson:2020vap,HegadeKR:2024agt,Katagiri:2024wbg}. Then, the analytic expressions for the electric-type and magnetic-type ``bare" tidal response functions are 
\begin{align}
\label{eq:tidalresponsefunctions}
{\cal F}_\ell^+\left(\omega\right)=&\frac{1}{2}{\cal C}^{2\ell+1}\tilde{F}^+\left(\omega\right),\\ 
{\cal F}_\ell^-\left(\omega\right)=&-\frac{\ell}{2\left(\ell+1\right)}{\cal C}^{2\ell+1}\tilde{F}^-\left(\omega\right),
\end{align}
where we have introduced a compactness,~${\cal C}:= M/r_0$, and 
\begin{widetext}
\begin{align}
    \tilde{F}^\pm \left(\omega\right):=&\frac{\mathbb{I}_{H/h}^{(0,0)}+\epsilon \mathbb{I}_{H/h}^{(1,0)}+\epsilon^2 \mathbb{I}_{H/h}^{(2,0)}+\omega M\left(\mathbb{I}_{H/h}^{(0,1)}+\epsilon \mathbb{I}_{H/h}^{(1,1)}+\epsilon^2 \mathbb{I}_{H/h}^{(2,1)}\right)}{\mathbb{E}_{H/h}^{(0,0)}+\epsilon \mathbb{E}_{H/h}^{(1,0)}+\epsilon^2 \mathbb{E}_{H/h}^{(2,0)}+\omega M\left(\mathbb{E}_{H/h}^{(0,1)}+\epsilon \mathbb{E}_{H/h}^{(1,1)}+\epsilon^2 \mathbb{E}_{H/h}^{(2,1)}\right) }.
\end{align}
\end{widetext}
Note that imposing an inner boundary condition fails to specify ${\cal F}_\ell^{\pm}$ uniquely due to the ambiguity of $P_{H/h}^{(i,j)}$ in Eqs.~\eqref{eq:H1i}--\eqref{eq:h2i}~\cite{HegadeKR:2024agt,Katagiri:2024wbg}.  

In the current perturbative framework, bare TLNs and TDNs are expanded in $\epsilon$:
\begin{align}
    \kappa_\ell^{\pm}\simeq& \kappa_\ell^{\pm(0)}+\epsilon \kappa_\ell^{\pm(1)}+\epsilon^2 \kappa_\ell^{\pm(2)},\\
    \nu_\ell^{\pm}\simeq& \nu_\ell^{\pm(0)}+\epsilon \nu_\ell^{\pm(1)}+\epsilon^2\nu_\ell^{\pm(2)}.
\end{align}
For a Schwarzschild BH, $\kappa_\ell^{\pm(0)}$ and $\nu_\ell^{\pm(0)}$ are known as~\cite{
Binnington:2009bb,LeTiec:2020spy,LeTiec:2020bos,Charalambous:2021mea,Chia:2020yla,Poisson:2020vap,Poisson:2021yau,Katagiri:2023yzm,Chakraborty:2023zed,Katagiri:2024wbg},
\begin{align}
    \kappa_\ell^{\pm(0)}=&0,\label{eq:GRTLNs}\\
    \nu_\ell^{\pm(0)}=&\frac{\left(\ell+2\right)!\ell! \left(\ell-1\right)!\left(\ell-2\right)!}{2\left(2\ell+1\right)!\left(2\ell-1\right)!}.\label{eq:GRTDNs}
\end{align}

\subsubsection{Canonical tidal response function}
We introduce CTRFs through the following two steps. The first is to determine particular solutions with the normalization originally proposed by Ref.~\cite{HegadeKR:2024agt} in the context of dynamical tides of neutron stars. The normalization provides a boundary condition for the particular solutions, which demands that {\it the asymptotic expansions of $P_{H}^{(i,j)}$~($P_{h}^{(i,j)}$, resp.) at large distances do not contain any terms proportional to $\bar{r}^{\ell}$ and $\bar{r}^{-\ell-1}$~($\bar{r}^{\ell+1}$ and $\bar{r}^{-\ell}$, resp.), where $\bar{r}$ is a harmonic radial coordinate}. Thus, one can systematically fix their functional form in a unified manner, allowing us to compare tidal response functions in various different setups in a meaningful way. We adopt the same boundary condition even for particular solutions of scalar-field or vector-field perturbations that often arise in extensions of GR with scalar or vector degrees of freedom. 

The second step is a redefinition of tidal moments with the remaining degrees of freedom to transform~\cite{Poisson:2020vap,Poisson:2021yau,Katagiri:2024wbg}:
\begin{align}
\bar{\mathbb{E}}_{H/h}^{(0,1)}=&\mathbb{E}_{H/h}^{(0,1)}-\xi_{H/h}^{(0,1)},\nonumber\\
\bar{\mathbb{E}}_{H/h}^{(1,1)}=&\mathbb{E}_{H/h}^{(1,1)}-\xi_{H/h}^{(1,1)},\\
\bar{\mathbb{E}}_{H/h}^{(2,1)}=&\mathbb{E}_{H/h}^{(2,1)}-\xi_{H/h}^{(2,1)},\nonumber
\end{align}
Here, $\xi_{H/h}^{(0,1)}$, $\xi_{H/h}^{(1,1)}$, and $\xi_{H/h}^{(2,1)}$ are arbitrary but not necessarily infinitesimal constants. Induced multipole moments remain invariant under this redefinition if and only if TDNs are also altered as
\begin{widetext}
\begin{align}
&\bar{\nu}_\ell^{\pm(0)}= \nu_\ell^{\pm (0)}-i \frac{\xi_{H/h}^{(0,1)}}{\mathbb{E}_{H/h}^{(0,0)}}\kappa_\ell^{\pm (0)},\\
    &\bar{\nu}_\ell^{\pm(1)}=\nu_\ell^{\pm (1)} +\frac{i}{\mathbb{E}_{H/h}^{(0,0)}} \left[ \left( \frac{\mathbb{E}_{H/h}^{(1,0)}}{\mathbb{E}_{H/h}^{(0,0)}}\kappa_\ell^{\pm (0)} -\kappa_\ell^{\pm (1)}  \right) \xi_{H/h}^{(0,1)}  -\kappa_\ell^{\pm (0)} \xi_{H/h}^{(1,1)}  \right], \\
    &\bar{\nu}_\ell^{\pm(2)}= \nu_\ell^{\pm (2)}\nonumber\\
    &-\frac{i}{\mathbb{E}_{H/h}^{(0,0)}} \left[ \left(\left\{ \left(\frac{\mathbb{E}_{H/h}^{(1,0)}}{\mathbb{E}_{H/h}^{(0,0)}}\right)^2  -\frac{\mathbb{E}_{H/h}^{(2,0)}}{\mathbb{E}_{H/h}^{(0,0)}}
 \right\}  \kappa_\ell^{\pm(0)}   - \frac{\mathbb{E}_{H/h}^{(1,0)}}{\mathbb{E}_{H/h}^{(0,0)}} \kappa_\ell^{\pm (1)} +\kappa_\ell^{\pm (2)}  \right) \xi_{H/h}^{(0,1)}  -\left( \frac{\mathbb{E}_{H/h}^{(1,0)}}{\mathbb{E}_{H/h}^{(0,0)}} \kappa_\ell^{\pm(0)} -\kappa_\ell^{\pm(1)}   \right) \xi_{H/h}^{(1,1)}+ \kappa_\ell^{\pm (0)}  \xi_{H/h}^{(2,1)} \right] .
\end{align}
\end{widetext}
Now, we choose $\xi_{H/h}^{(0,1)}=\mathbb{E}_{H/h}^{(0,1)}$, $\xi_{H/h}^{(1,1)}=\mathbb{E}_{H/h}^{(1,1)}$, and $\xi_{H/h}^{(2,1)}=\mathbb{E}_{H/h}^{(2,1)}$ so that {\it the contributions from the non-adiabatic regime in tidal moments are removed}. In the vacuum GR limit of~$\epsilon\to 0$, this recovers the redefinition performed in Refs.~\cite{Poisson:2020vap,Katagiri:2024wbg}.\footnote{One can shift even $\mathbb{E}_{H/h}^{(1,0)}$ and $\mathbb{E}_{H/h}^{(2,0)}$ to make them zero by redefinitions. In this work, we leave them to keep the framework simple considerably while being compatible with Refs.~\cite{Poisson:2020vap,Poisson:2021yau,Katagiri:2024wbg} in the vacuum GR limit.
} We thus define a CTRF,
\begin{align}
&\hat{\cal F}_\ell^{{\rm C} \pm}\left(\omega\right):=\kappa_\ell^{\pm(0)}+\epsilon \kappa_\ell^{\pm(1)}+\epsilon^2 \kappa_\ell^{\pm(2)}\label{eq:CTRFs}\\
&+i \omega M \left(\nu_\ell^{{\rm C}\pm(0)}+\epsilon \nu_\ell^{{\rm C}\pm(1)}+\epsilon^2 \nu_\ell^{{\rm C}\pm(2)}\right)+{\cal O}\left(\epsilon^3,\omega^2\right).\nonumber
\end{align}
with 
\begin{widetext}
\begin{align}
&\nu_\ell^{{\rm C}\pm(0)}:= \nu_\ell^{\pm (0)}-i \frac{\mathbb{E}_{H/h}^{(0,1)}}{\mathbb{E}_{H/h}^{(0,0)}}\kappa_\ell^{\pm (0)},\\
    &\nu_\ell^{{\rm C}{\pm(1)}}:=\nu_\ell^{\pm (1)} +\frac{i}{\mathbb{E}_{H/h}^{(0,0)}} \left[ \left( \frac{\mathbb{E}_{H/h}^{(1,0)}}{\mathbb{E}_{H/h}^{(0,0)}}\kappa_\ell^{\pm (0)} -\kappa_\ell^{\pm (1)}  \right) \mathbb{E}_{H/h}^{(0,1)}  -\kappa_\ell^{\pm (0)} \mathbb{E}_{H/h}^{(1,1)}  \right], \\
    &\nu_\ell^{{\rm C}\pm(2)}:= \nu_\ell^{\pm (2)}\nonumber\\
    &-\frac{i}{\mathbb{E}_{H/h}^{(0,0)}} \left[ \left(\left\{ \left(\frac{\mathbb{E}_{H/h}^{(1,0)}}{\mathbb{E}_{H/h}^{(0,0)}}\right)^2  -\frac{\mathbb{E}_{H/h}^{(2,0)}}{\mathbb{E}_{H/h}^{(0,0)}}
 \right\}  \kappa_\ell^{\pm(0)}   - \frac{\mathbb{E}_{H/h}^{(1,0)}}{\mathbb{E}_{H/h}^{(0,0)}} \kappa_\ell^{\pm(1)} +\kappa_\ell^{\pm (2)}  \right) \mathbb{E}_{H/h}^{(0,1)}  -\left( \frac{\mathbb{E}_{H/h}^{(1,0)}}{\mathbb{E}_{H/h}^{(0,0)}} \kappa_\ell^{\pm(0)} -\kappa_\ell^{\pm(1)}   \right) \mathbb{E}_{H/h}^{(1,1)}+ \kappa_\ell^{\pm (0)}  \mathbb{E}_{H/h}^{(2,1)} \right] .
\end{align}
\end{widetext}

\subsection{Comparison with an alternative approach}

Let us compare the approach to determine particular solutions in the current work to that in the context of dynamical tides in vacuum GR, proposed in Ref.~\cite{Katagiri:2024wbg}. The latter applies to particular solutions that appear in the low-frequency expansion of the linearized (vacuum) Einstein equations, leading to zero dynamical TLNs for a Schwarzschild BH in both even and odd sectors. This calibration allows us to quantify linear tidal responses, including up to dynamical regimes, of relativistic stars, such as neutron stars, within GR as the differences from the BH value in a unified and simple manner.

Following this spirit, one can set tidal response functions of any BHs to those in GR~(e.g., zero TLNs), even in beyond-vacuum GR scenarios, thereby simplifying the unified measurement of linear tidal responses of any non-BH objects as the difference from the BH values in a given theory. There are certain advantages. However, this approach privileges any BHs equally, resulting in multiple baselines depending on the underlying theory of gravity or the surrounding matter environments, despite the fact that neither is known {\it a priori}. This would, in turn, make comparisons of results difficult due to differences among numerous different theories or setups.

We find the CTRF more convenient and simpler because it privileges only the Schwarzschild BH in GR as the unique baseline. In other words, the CTRFs of any spherical objects quantify modification of the theory of gravity, differences in the equation of state, or other properties of compact objects and their surroundings, in terms of their difference from the value of the Schwarzschild BH within GR, in a unified manner. Note that the CTRF is not the unique way to set the Schwarzschild BH in GR as a baseline. We believe its construction is the simplest. Indeed, as will be seen in the next section, the CTRFs recover most of the known results for BHs in alternative theories of gravity, even though the tidal response functions in the literature were defined in a model-dependent and subjective manner.

\section{Applications}\label{Sec:Application}

\subsection{Theoretical predictions}
We compute the CTRFs for non-rotating BHs in Einstein-Maxwell, dynamical Chern-Simons~(dCS), Einstein-dilaton-Gauss-Bonnet~(EdGB) gravity, the braneworld scenario, and an effective field theory~(EFT) framework. These theories are precisely given in Appendix~\ref{Appendix:model}. 

The results are presented in Tables~\ref{table:TLNs} and~\ref{table:TDNs}. We recover the results for $\ell=2$ of GR~\cite{Binnington:2009bb,LeTiec:2020spy,LeTiec:2020bos,Charalambous:2021mea,Chia:2020yla,Poisson:2020vap,Poisson:2021yau,Katagiri:2024wbg}, Einstein-Maxwell~\cite{Cardoso:2017cfl,Poisson:2021yau}, dCS~\cite{Cardoso:2017cfl}, the braneworld BH~\cite{Tan:2020hog}, EFT framework~\cite{Cardoso:2018ptl} together with the parametrized formalism~\cite{Katagiri:2023umb}, while we find discrepancies for the braneworld BH in another analysis~\cite{Chakravarti:2018vlt}.  To the best of our knowledge, the BH TLNs in EdGB gravity and TDNs in non-GR theories are computed for the first time. 

We explain the causes for the disagreement with the literature~\cite{Chakravarti:2018vlt} for the braneworld BH in the following: First of all, the analysis in Ref.~\cite{Chakravarti:2018vlt} has a minor error in the derivation of perturbation equations~(see the details of Ref.~\cite{Tan:2020hog}), and moreover, the expression for TLNs given by Eq.~(41) in the literature, specifically $B/A$ therein, has no guarantee to capture the boundary condition at the BH horizon correctly, implying that the results in Ref.~\cite{Chakravarti:2018vlt} are incorrect. Then, Ref.~\cite{Tan:2020hog} corrected the error and obtained a different BH tidal deformability under some assumptions~(including the regularity of one of the homogeneous solutions at the horizon and the form of the source term in Eq.~(4.19)). Our approach rests on the consistent perturbative expansion of the perturbation equations under the unified definition, and then, derives the consistent results with Ref.~\cite{Tan:2020hog}, supporting the correctness of the analysis in the literature.

\subsection{Observational constraints}
One might be concerned about the potential misinterpretation of observational results arising from the identification of tidal response functions with tidal parameters incorporated in a PN gravitational waveform, adopted as an effective approach often used in previous studies, such as constraining neutron star radii and the equation of state~\cite{Flanagan:2007ix,LIGOScientific:2018hze,LIGOScientific:2018cki}. Here, we argue that the ambiguities highlighted in this work may not be the source of significant issues in the interpretation of current observations under this effective identification. More precisely, the current sensitivity is not sufficient to be concerned with the ambiguities. We demonstrate below that current observational data cannot impose meaningful constraints on the coupling constants of the theories considered in this work, regardless of the ambiguities, due to the large statistical uncertainties in their measurements. Third generation detectors will reduce measurement errors, but the ambiguities would become a major concern in the absence of a consistent framework that integrates tidal responses, binary evolution, and gravitational-wave emissions.

To see the above, we exploit the theory-agnostic constraints for binary-averaged quantities of TLNs and TDNs with compact binary merger events by Refs.~\cite{Narikawa:2021pak,Chia:2024bwc}. The binary tidal deformability for two bodies of mass $M_1$ and $M_2$~($M_1\ge M_2$) is defined by~\cite{Flanagan:2007ix,Hinderer:2007mb}
\begin{align}
    \tilde{\Lambda}:= \frac{16}{13} \frac{\left(M_1+12M_2\right)M_1^4\Lambda_1+\left(M_2+12 M_1\right)M_2^4 \Lambda_2}{\left(M_1+M_2\right)^5},
\end{align}
where $\Lambda_i$ is the tidal deformability parameter of the body of $M_i$, defined by $\Lambda_i := (2/3)(r_i/M_i)^5\kappa_{i,2}^+$ with the radius~$r_i$ and the electric-type quadrupolar TLN~$\kappa_{i,2}^+$. 

Assuming that the two objects are non-spinning BHs in theories that slightly deviate from vacuum GR, the above~$\tilde{\Lambda}$ is parametrized by $M_1$, $M_2$, as well as the coupling constant. We translate the $90\%$ symmetric credible ranges of $\tilde{\Lambda}$ for GW151226, GW170608, GW190707, GW190720, GW190728, GW190924 by Ref.~\cite{Narikawa:2021pak} into the bounds for the coupling constants in EdGB, braneworld scenario, and EFT~($\epsilon_1$) with Table~\ref{table:TLNs} by {\it identifying} $\kappa_{i,2}^+$ with the TLNs from the CTRF~\eqref{eq:CTRFs}, although they are not necessarily the same. The magnitude of the ambiguities is at most of order of the coupling constants. We found that the absolute values of the resulting bounds for those theories are ten times larger than the upper bounds for which the perturbative expansion of the CTRFs in the coupling constants are valid. This holds true even when maximizing the values of $\kappa_{i,2}^+$ by exploiting the ambiguities. We thus conclude that the current theory-agnostic observational constraints for $\tilde{\Lambda}$ cannot set meaningful constraints on the small coupling constants, regardless of the ambiguities.

We next explain the TDN case.
The mass-weighted combination of (spin-independent) tidal dissipation parameters is defined as~\cite{Chia:2024bwc}
\begin{align}
    {\cal H}_0:= \frac{M_1^4 H_{1\omega} +M_2^4 H_{2\omega}}{\left(M_1+M_2\right)^4},
\end{align}
where $H_{i \omega}$ is related to the TDNs of the body of mass~$M_i$ via $H_{i \omega}=(2/3)(r_i/M_i)^4 \nu_{i,2}^+$. In the same manner as the TLNs, we translate the $90\%$ credible intervals for ${\cal H}_0$ of GW190514, GW170608, GW170814, GW190708, GW191216, GW200202, GW200311 by Ref.~\cite{Chia:2024bwc} into the bounds for the coupling constants in EdGB, braneworld scenario, and EFT~($\epsilon_1$) with Table~\ref{table:TDNs}. The magnitude of the ambiguities is at most of order of the coupling constants. Again, the absolute values of the bounds for those theories are ten times larger than the upper bounds for which the perturbative expansion of the CTRFs in the coupling constants remains valid. This indicates that the current theory-agnostic observational constraints for ${\cal H}_0$ cannot set meaningful constraints on the small coupling constants, regardless of the ambiguities.

\section{Conclusion and discussion}\label{Sec:Conclusion}
We have proposed a {\it canonical tidal response function}~(CTRF) that allows us to compare tidal response functions for any compact objects beyond vacuum GR under the unified definition. As a demonstration, we present BH CTRFs in various alternative theories of gravity in Tables~\ref{table:TLNs} and~\ref{table:TDNs}. These values allow us to define tidal deformabilities of relativistic stars, such as neutron stars, as the difference from BH values in each non-GR theory in a unified manner.

Our CTRFs recover most of the known results for BHs in alternative theories of gravity even though the tidal response functions in the literature were defined in a model-dependent subjective way, suggesting that the CTRF is the simplest and natural unified definition. Our proposal not only reorganizes previous results in a unified manner but also allows us to compare tidal responses across different setups beyond vacuum GR in a meaningful way in the future. 

We argued that the ambiguities highlighted in this work may not be the source of significant issues in the interpretation of current observations under the effective identification of tidal response functions with tidal parameters incorporated in a PN gravitational waveform. More precisely, the current sensitivity is not sufficient to be concerned with the ambiguities. We demonstrated that current observations cannot impose meaningful bounds on the small coupling constants of the theories considered in this work, regardless of the ambiguities. To see this, we translated the theory-agnostic observational constraints for the binary-averaged tidal deformabilities of compact binary merger events by Refs.~\cite{Narikawa:2021pak,Chia:2024bwc} into bounds for the coupling constants of EdGB, the braneworld scenario, and EFT~($\epsilon_1$) with our theoretical predictions. The absolute values of the resulting bounds are ten times larger than the upper bound for which the perturbative expansion of the theories about the coupling constants remain valid. This remains true even when the tidal response functions are maximized by exploiting the ambiguities. Third generation detectors will improve sensitivities, and then the ambiguities would become a major concern in the absence of a consistent framework that integrates tidal responses, binary evolution, and gravitational-wave emissions.

Let us comment on the limitations of our framework. Our approach relies on an approximate matching scheme of a tidally deformed body metric with a PN metric at the leading Newtonian order in vacuum GR. We then assume that a Newtonian potential can approximate the tidal potential of a BH. This remains valid for a binary system with sufficiently large separations compared to the radii of the binary constituents. In the vacuum GR limit, our framework recovered the vanishing BH TLNs derived in a fully relativistic setup in Ref.~\cite{Poisson:2021yau}. Additionally, the harmonic coordinate system used in this work is not exactly harmonic in tidally perturbed spacetimes but should still be a good approximation in a certain inspiral stage. It would be beneficial for a better understanding to compare the full PN order results derived in a manner similar to Ref.~\cite{Poisson:2021yau} with our Newtonian order results.

Let us mention an alternative approach to determine particular solutions in Ref.~\cite{Barura:2024uog}. The analysis in the literature decomposes an integration constant into tidal and response pieces in a different way from ours and Ref.~\cite{Katagiri:2023umb}, leading to apparently different TLNs. The presence of a homogeneous piece in particular solutions explains their findings of a cancellation between the contributions from the growing and decaying modes in the large-distance expansion of the perturbation. The analytic continuation of $\ell$ argued in the literature is not essential. Additionally, the formalism presented in Appendix~B of Ref.~\cite{Barura:2024uog} appears not to recover the results of Ref.~\cite{Cardoso:2018ptl}~(note the difference of the numerical coefficient in the definition of TLNs from the original parametrized formalism~\cite{Katagiri:2023umb}). Again, this arises from the different decomposition but that does not immediately indicate a disagreement in the predictions of observables.

We end by presenting a few possible avenues for future work.
One possible extension of our current work is to reformulate a parametrized BH TLN formalism~\cite{Katagiri:2023umb}. The original work~\cite{Katagiri:2023umb} does not take into account the ambiguity argued here, potentially leading to apparent discrepancies~\cite{Barura:2024uog}. Second, an extension of our framework to rotating backgrounds is crucial to accurately modeling tidal effects of astrophysical BHs. Last, one needs to study how exactly the TLNs/TDNs defined here enter in gravitational waveforms from compact binary inspirals and carry out more accurate analysis for constraining coupling parameters in alternative theories of gravity from observational data.

\acknowledgments
We acknowledge support by VILLUM Foundation (grant no.\ VIL37766) and the DNRF Chair program (grant no.\ DNRF162) by the Danish National Research Foundation.
V.C.\ acknowledges financial support provided under the European Union’s H2020 ERC Advanced Grant “Black holes: gravitational engines of discovery” grant agreement no.\ Gravitas–101052587. 
Views and opinions expressed are however those of the author only and do not necessarily reflect those of the European Union or the European Research Council. Neither the European Union nor the granting authority can be held responsible for them.
This project has received funding from the European Union's Horizon 2020 research and innovation programme under the Marie Sk{\l}odowska-Curie grant agreement No 101007855 and No 101131233.
K.Y. acknowledges support from NSF Grant No.~PHY-2207349, No.~PHY-2309066, No.~PHYS-2339969, and the Owens Family Foundation.

\appendix

\section{Static-binary model: Majumdar-Papapetrou spacetime}\label{sec:MP}
A Majumdar-Papapetrou metric is a multiple BH solution in Einstein-Maxwell theory, whose geometry is interpreted as a composite system of extremal Reissner-Nordstr{\"o}m BHs in equilibrium under their gravitational and electric forces~\cite{PhysRev.72.390,2b9b5749-2539-3ff6-8e72-d50ecc64b745,Hartle:1972ya}. Here, we consider a dihole system with two BHs of mass~$M_1$ and $M_2$. The line element in the static~$(t,r,\theta,\varphi)$ coordinates associated with the BH of mass~$M_1$ is given by
\begin{align}
\label{eq:MPmetric}
    ds^2=-N^2 dt^2+\frac{dr^2}{N^2}+\left(1-\frac{M_1}{r}\right)^2\frac{r^2}{N^2}d\Omega^2,
\end{align}
where  
\begin{align}
\label{eq:lapce}
    \frac{1}{N}=\frac{1}{1-M_1/r}+\frac{M_2}{\sqrt{\left(r-M_1-2b\cos\theta\right)^2+4b^2\sin^2\theta}}\,,
\end{align}
with $b$ representing the separation between the two BHs.
When $M_2=0$, Eq.~\eqref{eq:MPmetric} reduces to an extremal Reissner-Nordstr{\"o}m metric. 

Now, we consider the above metric in the so-called near zone~$r<b$. Then, Eq.~\eqref{eq:lapce} is cast into  
\begin{align}
    \label{eq:lapceinY}
    \frac{1}{N}=&\frac{1}{1-M_1/r}\nonumber \\
    &-\sum_{\ell=0}^\infty \left(\frac{4\pi}{2\ell+1}\right)^{1/2}\frac{M_2}{2b}\left(\frac{r-M_1}{2b}\right)^\ell Y_\ell,
\end{align}
where $Y_{\ell}=Y_{\ell 0}(\theta)$. The second line is responsible for anisotropic deformation of the extremal Reissner-Nordstr{\"o}m BH metric, induced by the presence of $M_2$, and hence, it is interpreted as tidal interaction terms between the two BHs. For simplicity, we expand the $tt$ component in terms of the mass ratio~$\eta(:=M_2/M_1)$, obtaining the leading quadrupole order~$\ell=2$:
\begin{align}
    -N^2\big|_{\ell=2}=&-\left(1-\frac{M_1}{r}\right)^2\\
    &-\left(\frac{4\pi}{5}\right)^{1/2}\frac{M_1 \eta}{4b^3} \frac{\left(r-M_1\right)^5}{r^3}Y_2 +{\cal O}\left(\eta^2\right).\nonumber
\end{align}
The tidal interaction term consists of the mixing between the contributions from both BHs. This shows that there is no unambiguous way to decompose it into an external tidal piece for $M_1$ by $M_2$ and an induced response piece of $M_1$. The limit of $M_1\to 0$ corresponds to a pure contribution from $M_2$, and then, one can read off the quadrupolar electric-type tidal moment~$(4\pi/5)^{1/2}M_2/(8b^3) Y_2$.

To be clear, we provide the effective gravitational potential,~$U_{\rm eff}:=(1+g_{tt})/2$, in $M\ll r\ll b$,
\begin{align}
  U_{\rm eff}=&\frac{M_1}{r}-\frac{M_1^2}{2r^2}\nonumber \\
  &-\left(\frac{4\pi}{5}\right)^{1/2}\frac{M_1 \eta}{8b^3}r^2\left[1+{\cal O}\left(M_1/r\right)\right]  Y_2+{\cal O}\left(\eta^2\right).
\end{align}
Note that the first line terminates at ${\cal O}(r^{-2})$, while the series in the second line terminates at ${\cal O}(r^{-3})$ for any $\ell$. The second line is supposed to be responsible for the tidal deformation of $M_1$ but, again, there is no unambiguous way to decompose the series into external tidal and induced response pieces. 

Within the BH perturbation theory on the extremal Reissner-Nordrstr{\"o}m spacetime, one can quantify the tidal deformation of $M_1$ with the CTRF~\eqref{eq:CTRFs} in a unified manner. The gravitational perturbation is coupled with the Maxwell-field perturbation~(see Eqs.~\eqref{eqsforHURN} and~\eqref{eqsforhuRN} with $M=Q$). The coupled system is solvable in closed form. With the CTRF~\eqref{eq:CTRFs}, one can show the vanishing of TLNs of the extremal Reissner-Nordstr{\"o}m BH under the assumption of the absence of an external electromagnetic tidal field, following Ref.~\cite{Cardoso:2017cfl}.

\section{Low-frequency expansion of linear perturbations in vacuum GR}\label{App:GRsolution}
We here provide the general solutions and their properties of linear perturbations on a Schwarzschild spacetime within low-frequency expansions to first order. Section~\ref{sec:MetricPer} is devoted to metric perturbations, and then, vector-field and massless-scalar-field perturbations are outlined in Secs.~\ref{sec:VectorPer} and~\ref{sec:ScalarPer}. We last present the general solutions of Zerilli-Regge-Wheeler equations~\cite{PhysRev.108.1063,PhysRevD.2.2141} in Sec.~\ref{sec:RWZformalism}.

\subsection{Metric perturbation}\label{sec:MetricPer}
Consider a Schwarzschild background in $(t,r,\theta,\varphi)$ coordinates:
\begin{align}
    g_{\mu \nu}^{(0)} dx^\mu dx^\nu =-fdt^2+f^{-1}dr^2+ r^2d\Omega^2,
\end{align}
where $f:=1-2M/r$ and $d\Omega^2:=d\theta^2+\sin^2\theta d\varphi^2$. A metric perturbation~$h_{\mu\nu}$ on the Schwarzschild spacetime is decomposed into even- and odd-parity sectors, subject to the parity transformation~${\bf P}:(\theta,\varphi)\to (\pi-\theta,\varphi+\pi)$. In Fourier domain, the components of each sector with a spherical harmonic decomposition in the Regge-Wheeler gauge~\cite{PhysRev.108.1063} are 
\begin{widetext}
\begin{align}\label{eq:Heven}
    \left(h_{\ell m}^{\rm (even)}\right)_{\mu\nu}=\begin{pmatrix}
fH^{\ell m}Y_{\ell m}e^{-i \omega t} & H_1^{\ell m} Y_{\ell m}e^{-i \omega t} & 0 &0\\
H_1^{\ell m} Y_{\ell m}e^{-i \omega t} & f^{-1}H_2^{\ell m} Y_{\ell m}e^{-i \omega t} &  0 &0\\
0 & 0 & r^2K^{\ell m} Y_{\ell m}e^{-i \omega t}&0\\
0 & 0 & 0  &r^2\sin^2\theta K^{\ell m} Y_{\ell m}e^{-i \omega t}\\
\end{pmatrix},
\end{align}
and
\begin{align}\label{eq:hodd}
    \left(h_{\ell m}^{\rm (odd)}\right)_{\mu\nu}=\begin{pmatrix}
0 &0 & h^{\ell m}  S_\theta^{\ell m}e^{-i \omega t} &h^{\ell m} S_\varphi^{\ell m}e^{-i \omega t} \\
0 & 0 &  h_1^{\ell m} S_\theta^{\ell m}e^{-i \omega t}  & h_1^{\ell m} S_\varphi^{\ell m}e^{-i \omega t}\\
h^{\ell m} S_\theta^{\ell m}e^{-i \omega t}  & h_1^{\ell m} S_\theta^{\ell m}e^{-i \omega t} &0 &0\\
h^{\ell m} S_\varphi^{\ell m}e^{-i \omega t}  & h_1^{\ell m} S_\varphi^{\ell m}e^{-i \omega t}& 0  &0\\
\end{pmatrix},
\end{align}
\end{widetext}
where $H_i^{\ell m}=H_i^{\ell m}(r)$, $K^{\ell m}=K^{\ell m}(r)$, $h^{\ell m}=h^{\ell m}(r)$; $Y_{\ell m}=Y_{\ell m}(\theta,\varphi)$ is a scalar spherical harmonic; $(S_\theta^{\ell m} ,S_\varphi^{\ell m} )=(-\partial_\varphi Y_{\ell m} /\sin\theta,\sin\theta \partial_\theta Y_{\ell m} )$.

The linearized Einstein equations in vacuum, $\delta G_{\mu \nu}=0$, can be reduced to equations for $H^{\ell m}$ and $h^{\ell m}$ in each sector~\cite{Katagiri:2024wbg}. We henceforth omit the superscript~$\ell m$. Assuming low frequency, we expand $H$ and $h$ in $\omega M$ up to first order:
\begin{align}
    H=&H^{(0,0)}+\omega M H^{(0,1)}+{\cal O}\left(\omega^2 M^2\right),\\
    h=&h^{(0,0)}+\omega M h^{(0,1)}+{\cal O}\left(\omega^2 M^2\right).
\end{align}
Here, the superscript~$(0,i)$ means the $i$th order of the low-frequency expansion. Then, the resulting equations up to ${\cal O}(\omega M)$ take the forms,
\be
{\cal L}_H\left[H^{(0,i)}\right]=0\,,
 {\cal L}_h\left[h^{(0,i)}\right]=0,\label{eqforHh}
\ee
with
\begin{align}
    {\cal L}_H:=& \frac{d^2}{dr^2}+\frac{2\left(r-M\right)}{r^2 f}\frac{d}{dr}-\frac{1}{r^2f}\left[\ell\left(\ell+1\right)+\frac{4M^2}{r^2f}\right],\label{eq:LHGR}\\
    {\cal L}_h:=& \frac{d^2}{dr^2}-\frac{1}{r^2f}\left[\ell\left(\ell+1\right)-\frac{4M}{r}\right].\label{eq:LhGR}
\end{align}

The general solutions of Eq.~\eqref{eqforHh} take the form,
\begin{equation}
H^{(0,i)}=\mathbb{E}_H^{(0,i)}  H_T +\mathbb{I}_H^{(0,i)} H_R ,\, h^{(0,i)}  =\mathbb{E}_h^{(0,i)}  h_T +\mathbb{I}_h^{(0,i)}  h_R,\label{eqs:H0ih0i} 
\end{equation}
which is the same form as Eq.~\eqref{eq:HhGR}. Here, $\mathbb{E}_{H/h}^{(0,i)} $ and $\mathbb{I}_{H/h}^{(0,i)}$ are integration constants, with
\begin{align}
    H_T  :=&f\left(\frac{r}{M}\right)^{\ell}\!~_2F_1\left(-\ell+2,-\ell;-2\ell;2M/r\right),\label{eq:HT}\\
    H_R   :=&f \left(\frac{M}{r}\right)^{\ell+1}\!~_2F_1\left(\ell+1,\ell+3;2\ell+2;2M/r\right)\label{eq:HR}, 
\end{align}
and
\begin{align}
    h_T   :=&\left(\frac{r}{M}\right)^{\ell+1}\!~_2F_1\left(-\ell+1,-\ell-2;-2\ell;2M/r\right),\label{eq:hT}\\
    h_R   :=&\left(\frac{M}{r}\right)^{\ell}\!~_2F_1\left(\ell-1,\ell+2;2\ell+2;2M/r\right),\label{eq:hR} 
\end{align}
where~$_2F_1(a,b;c;2M/r)$ are Gaussian hypergeometric functions. For the Schwarzschild BH, the imposition of an ingoing-wave condition at the horizon leads to the vanishing TLNs~\eqref{eq:GRTLNs} and the finite TDNs~\eqref{eq:GRTDNs}, as well as the relation~\cite{Katagiri:2024wbg},
\begin{align}
    \mathbb{E}_H^{(0,1)}=&4i \frac{\ell\left(\ell+1\right)\left(\psi\left(\ell\right)+\gamma\right)-\ell^2-1}{\ell\left(\ell+1\right)}  \mathbb{E}_H^{(0,0)},\label{eq:E0HE1H}\\
    \mathbb{E}_h^{(0,1)}=&4i\left(\gamma+\psi\left(\ell\right)+\frac{\ell^3-3\ell-1}{\left(\ell+2\right)\left(\ell+1\right)\ell\left(\ell-1\right)}\right)\mathbb{E}_h^{(0,0)},\label{eq:E0hE1h}
\end{align}
where $\psi(\ell)$ and $\gamma(\simeq 0.57721)$ are a digamma function and Euler's constant, respectively.

The Wronskian and the asymptotic behavior of the perturbation fields are given as follows.
Defining the Wronskian~$W[g_1,g_2]:=g_1 \frac{d}{dr}g_2-g_2 \frac{d}{dr}g_1$, we have
\begin{align}
    W\left[H_T ,H_R \right]=&-\frac{2\ell+1}{Mf}\left(\frac{M}{r}\right)^2=:W_{H},\label{eq:WH}\\
   W\left[h_T ,h_R \right]=&-\frac{2\ell+1}{M}=:W_{h}.\label{eq:Wh}
\end{align}
The functions~$H_{T/R}$ and $h_{T/R}$ behave asymptotically as
\begin{align}
    H_T \sim &{\cal O}\left(f\right),~   H_R \sim {\cal O}\left(f^{-1},\ln f\right),\\ 
      h_T \sim&{\cal O}\left(f\right),~      h_R \sim {\cal O}\left(\ln f\right),
\end{align}
near the BH horizon, and
\begin{align}
     H_T \big|_{r\gg M}=&\left(\frac{r}{M}\right)^\ell\left[1+{\cal O}\left(M/r\right)\right],\\ 
     H_R \big|_{r\gg M}=&\left(\frac{M}{r}\right)^{\ell+1}\left[1+{\cal O}\left(M/r\right)\right],\\ 
      h_T \big|_{r\gg M}=&\left(\frac{r}{M}\right)^{\ell+1}\left[1+{\cal O}\left(M/r\right)\right],\\ 
      h_R \big|_{r\gg M}=&\left(\frac{M}{r}\right)^{\ell}\left[1+{\cal O}\left(M/r\right)\right],
\end{align}
at large distances.

\subsection{Vector-field perturbation}\label{sec:VectorPer}
Consider a vector-field perturbation $\delta A^\mu$ satisfying
\begin{align}
    \nabla_\nu \delta F^{\mu\nu}=0, 
\end{align}
with the field strength perturbation given by~$\delta F_{\mu \nu}:=\partial_\mu \delta A_\nu -\partial_\nu \delta A_\mu$. The components of $\delta A_\mu$ with a harmonic decomposition in Fourier domain are given by
\begin{align}
\label{deltaA}
    \left(\delta A_{\ell m}\right)_\mu =\begin{pmatrix}
    U^{\ell m} Y_{\ell m} e^{-i\omega t}\\
   a_r^{\ell m} Y_{\ell m} e^{-i\omega t}\\
     k^{\ell m} \partial_\theta Y_{\ell m} e^{-i\omega t}+\dfrac{u^{\ell m} }{\sin\theta} \partial_\varphi Y_{\ell m}  e^{-i\omega t}\\
      k^{\ell m} \partial_\varphi Y_{\ell m} -u^{\ell m} \sin\theta \partial_\theta Y_{\ell m} e^{-i\omega t}
     \end{pmatrix},
\end{align}
where $U^{\ell m}= U^{\ell m} (r)$,  $a_r^{\ell m}=a_r^{\ell m} (r)$, $k^{\ell m}=k^{\ell m} (r)$, and $u^{\ell m}=u^{\ell m} (r)$. The linearized Maxwell equations can be reduced to two equations for $U^{\ell m}$ and $u^{\ell m}$. Hereafter, we drop $\ell m$. Within the low-frequency expansion, $U$ and $u$ are expanded as
\begin{align}
    U=&U^{(0,0)}+\omega M U^{(0,1)} +{\cal O}\left(\omega^2 M^2\right),\\
     u=&u^{(0,0)}+\omega M u^{(0,1)} +{\cal O}\left(\omega^2 M^2\right).
\end{align}
The perturbation equations then take the forms of a homogeneous equation up to first order in $\omega M$,
\begin{align}
    {\cal L}_U\left[U^{(0,i)}\right]  =&0,~
     {\cal L}_u \left[u^{(0,i)}\right] =0,\label{eqforUu} 
\end{align}
where
\begin{align}
    {\cal L}_U:=& \frac{d^2}{dr^2}+\frac{2}{r}\frac{d}{dr}-\frac{\ell\left(\ell+1\right)}{r^2f} ,\label{eq:LUGR}\\
    {\cal L}_u:=& \frac{d^2}{dr^2}+\frac{2M}{r^2 f}\frac{d}{dr}  -\frac{\ell\left(\ell+1\right)}{r^2f}\label{eq:LuGR}.
\end{align}

Equation~\eqref{eqforUu} is solved as
\begin{equation}
U^{(0,i)}=\mathbb{E}_U^{(0,i)} U_T +\mathbb{I}_U^{(0,i)} U_R,\,u^{(0,i)}  =\mathbb{E}_u^{(0,i)} u_T +\mathbb{I}_u^{(0,i)} u_R\,,\label{eq:U0iu0i}
\end{equation}
where $\mathbb{E}_{U/u}^{(0,i)}$ and $\mathbb{I}_{U/u}^{(0,i)}$ are integration constants. We also introduced
\begin{align}
    U_T:=&\left(\frac{r}{M}\right)^{\ell}\!~_2F_1\left(-\ell,-\ell-1;-2\ell;2M/r\right),\label{eq:UT}\\
    U_R:=&\left(\frac{M}{r}\right)^{\ell+1}\!~_2F_1\left(\ell+1,\ell;2\ell+2;2M/r\right) ,\label{eq:UR} 
\end{align}
and
\begin{align}
    u_T:=&\left(\frac{r}{M}\right)^{\ell+1}\!~_2F_1\left(-\ell-1,-\ell+1;-2\ell;2M/r\right),\label{eq:uT}\\
    u_R:=& \left(\frac{M}{r}\right)^{\ell}\!~_2F_1\left(\ell,\ell+2;2\ell+2;2M/r\right).\label{eq:uR}
\end{align}
Then, we have the Wronskian,
\begin{align}
    W\left[U_T,U_R\right]=&-\frac{2\ell+1}{r^2}M=:W_{U},\label{eq:WU}\\
   W\left[u_T,u_R\right]=&-\frac{2\ell+1}{Mf}=:W_{u}.\label{eq:Wu}
\end{align}
The asymptotic behaviors of $U_{T/R}$ and $u_{T/R}$ are 
\begin{eqnarray}
U_T &\sim &{\cal O}\left(f\right),\,U_R\sim {\cal O}\left(f\ln f\right),\\ 
u_T &\sim &{\cal O}\left(f^0\right),\,
u_R\sim {\cal O}\left(\ln f\right),
\end{eqnarray}
near the BH horizon, and
\begin{align}
     U_T\big|_{r\gg M}=&\left(\frac{r}{M}\right)^{\ell}\left[1+{\cal O}\left(M/r\right)\right],\\ 
     U_R\big|_{r\gg M}=&\left(\frac{M}{r}\right)^{\ell+1}\left[1+{\cal O}\left(M/r\right)\right],\\ 
      u_T\big|_{r\gg M}=&\left(\frac{r}{M}\right)^{\ell+1}\left[1+{\cal O}\left(M/r\right)\right],\\ 
      u_R\big|_{r\gg M}=&\left(\frac{M}{r}\right)^{\ell}\left[1+{\cal O}\left(M/r\right)\right],
\end{align}
at large distances.

\subsection{Massless-scalar-field perturbation}\label{sec:ScalarPer}
Consider now a perturbation to a massless Klein-Gordon field
\begin{align}
    \Box \delta \Phi\left(t,r,\theta,\varphi\right)=0.
\end{align}
We decompose $\delta \Phi$ in spherical harmonics as
\begin{align}
\delta \Phi(t,r,\theta,\varphi)=\frac{\phi(r)}{r}Y_{\ell m}e^{-i \omega t},\label{deltaPhi}
\end{align}
and expand $\phi$ in terms of $\omega M$, 
\begin{align}
    \phi=\phi^{(0,0)}+\omega M \phi^{(0,1)} +{\cal O}\left(\omega^2 M^2\right).
\end{align}
The scalar field~$\phi^{(0,i)}$ satisfies
\begin{align}
    {\cal L}_\phi\left[\phi^{(0,i)}\right]=0, \label{eqforphi}
\end{align}
where
\begin{align}
    {\cal L}_\phi:=\frac{d^2}{dr^2}+\frac{2M}{r^2f}\frac{d}{dr}-\frac{1}{r^2f}\left[\ell\left(\ell+1\right)+\frac{2M}{r}\right].\label{eq:LphiGR}
\end{align}

Equation~\eqref{eqforphi} admits the general solution,
\begin{align}
    \phi^{(0,i)}=&\mathbb{E}_\phi^{(0,i)} \phi_T +\mathbb{I}_\phi^{(0,i)} \phi_R,
\end{align}
where $\mathbb{E}_\phi^{(0,i)}$ and $\mathbb{I}_\phi^{(0,i)}$ are integration constants. We also introduced
\begin{align}
    \phi_T:=&\left(\frac{r}{M}\right)^{\ell+1}~_2F_1\left(-\ell,-\ell;-2\ell;2M/r\right),\label{eq:phiT}\\
    \phi_R:=&\left(\frac{M}{r}\right)^{\ell}~_2F_1\left(\ell+1,\ell+1;2\ell+2;2M/r\right). \label{eq:phiR}
\end{align}
The Wronskian is
\begin{align}
    W\left[\phi_T,\phi_R\right]=&-\frac{2\ell+1}{M f}=:W_{\phi}.\label{eq:Wphi}
\end{align}
The asymptotic behaviors of $\phi_{T/R}$ are 
\begin{equation}
\phi_T \sim {\cal O}\left(f^0\right),\,\phi_R \sim {\cal O}\left(\ln f\right),
\end{equation}
near the BH horizon, and
\begin{align}
     \phi_T\big|_{r\gg M}=&\left(\frac{r}{M}\right)^{\ell+1}\left[1+{\cal O}\left(M/r\right)\right],\\ 
     \phi_R\big|_{r\gg M}=&\left(\frac{M}{r}\right)^{\ell}\left[1+{\cal O}\left(M/r\right)\right],
\end{align}
at large distances.

\subsection{Regge-Wheeler/Zerilli formalism}\label{sec:RWZformalism}
The Zerilli-Regge-Wheeler equations for the metric perturbations are given by~\cite{PhysRev.108.1063,PhysRevD.2.2141,Moncrief:1974am}
\begin{equation}
\label{ZRWeqs}
  f \frac{d}{dr}\left( f\frac{d\chi^{\pm}_\ell}{dr}\right)+\left(\omega^2-fV_\ell^{\pm} \right)\chi_\ell^{\pm}=0,
\end{equation}
where 
\begin{align}
    V_{\ell}^+:=&\frac{36\lambda M^2r+6\lambda^2M r^2+\lambda^2\left(\lambda+2\right)r^3+72M^3}{r^3\left(\lambda r+6M\right)^2},\\
V_{\ell}^-:=&\frac{\ell\left(\ell+1\right)}{r^2}-\frac{6 M}{r^3},
\end{align}
with $\lambda:=\ell^2+\ell-2$. We expand the master functions $\chi_\ell^\pm$ as $\chi_\ell^\pm =\chi_\ell ^{\pm(0,0)}+\omega M \chi_\ell^{\pm(0,1)}+{\cal O}(\omega^2 M^2)$. Then, Eq.~\eqref{ZRWeqs} reduces to
\begin{equation}
\label{staticZRWeqs}
 \frac{d}{dr}\left( f\frac{d\chi^{\pm(0,i)}_\ell}{dr}\right)-V_\ell^{\pm}\chi_\ell^{\pm(0,i)}=0.
\end{equation}

The general solution of Eq.~\eqref{staticZRWeqs} can be written in the form,
\begin{align}
\label{eq:generalPhiZRW}
    \chi_\ell^{\pm(0,i)}=\mathbb{E}_\chi^{\pm(0,i)} \chi_T^{\pm}+\mathbb{I}_\chi^{\pm(0,i)} \chi_R^{\pm},
\end{align}
where $\mathbb{E}_\chi^{\pm(0,i)}$ and $\mathbb{I}_\chi^{\pm(0,i)}$  are integration constants, and 
\begin{widetext}
\begin{align}
    \chi_T^+:=&\frac{12}{\lambda\left(\lambda+2\right)}\left(\frac{r}{M}\right)^{\ell+1}\left[\left\{\frac{\lambda\left(\lambda+2\right)}{6}+2f\frac{M}{r}\left(\frac{\lambda r +12M}{\lambda r+6M}+\ell\right)\right\}~\!_2F_1\left(-\ell-2,-\ell+2;-2\ell;2M/r\right) \right.\nonumber\\
    &\quad\quad\quad\quad\quad\quad\quad\quad \left.+2f\frac{\ell^2-4}{\ell}\left(\frac{M}{r} \right)^2\!~_2F_1\left(-\ell-1,-\ell+3;-2\ell+1,2M/r\right)\right],\nonumber\\
\end{align}
\begin{align}
     \chi_R^+:=&\frac{3}{\lambda\left(\lambda+2\right)}\left(\frac{M}{r}\right)^\ell\left[\left\{\frac{\lambda\left(\lambda+2\right)}{3}+4f\frac{M}{r}\left(\frac{6M}{\lambda r+6M}-\ell\right)\right\} ~\!_2F_1\left(\ell-1,\ell+3;2\ell+2;2M/r\right) \right.\nonumber\\
    &\quad\quad\quad\quad\quad\quad\quad\quad \left.-4f\frac{\left(\ell+3\right)\left(\ell-1\right)}{\ell+1}\left(\frac{M}{r}\right)^2 \!~_2F_1\left(\ell,\ell+4;2\ell+3;2M/r\right)\right],\nonumber
\end{align}
and
\begin{align}
      \chi_T^{-}:=\left(\frac{r}{M}\right)^{\ell+1}\!_2F_1\left(-\ell-2,-\ell+2;-2\ell;2M/r\right),~~\chi_R^{-}:=\left(\frac{M}{r}\right)^\ell \!_2F_1\left(\ell-1,\ell+3;2\ell+2;2M/r\right).
\end{align}
\end{widetext}
The Wronskian is 
\begin{align}
    W\left[\chi_T,\chi_R\right]=&-\frac{2\ell+1}{M f}=:W_{\chi}.
\end{align}
The asymptotic behaviors of $\chi_{T/R}^\pm$ are
\begin{equation}
\chi_T^\pm\sim {\cal O}\left(f^0\right),\,
\chi_R^\pm \sim {\cal O}\left(\ln f\right),
\end{equation}
near the BH horizon, and
\begin{align}
     \chi_T^\pm \big|_{r\gg M}=&\left(\frac{r}{M}\right)^{\ell+1}\left[1+{\cal O}\left(M/r\right)\right],\\ 
     \chi_R^\pm \big|_{r\gg M}=&\left(\frac{M}{r}\right)^{\ell}\left[1+{\cal O}\left(M/r\right)\right],
\end{align}
at large distances.

\section{Tidal perturbations in particular models}\label{Appendix:model}
In this appendix, we provide details for tidal perturbations on a spherically symmetric background in particular models within two-parameter perturbative expansions of frequency and a coupling constant controlling the deviation from vacuum GR. The low-frequency expansion is performed in the manner reviewed in Appendix~\ref{App:GRsolution}. We also explain how to determine CTRFs~\eqref{eq:CTRFs} for BH backgrounds.

\subsection{Einstein-Maxwell theory}
The standard theory coupled with gravity and electromagnetic field is Einstein-Maxwell theory. The field equations are given by
\begin{align}
    G_{\mu \nu}&=8 \pi T_{\mu \nu}, \\
    \nabla_{\mu} F^{\mu \nu }&=0,
\end{align}
where $F_{\mu \nu}=\partial_\mu A_{\nu}-\partial_\nu A_{\mu}$ and
\begin{align}
    T_{\mu \nu}=\frac{1}{4 \pi}\left(F_{\mu \rho} F^\rho_{~\nu}-\frac{g_{\mu \nu}}{4} F_{\rho\sigma} F^{\rho \sigma} \right).
\end{align}
One can find an exact static and spherically symmetric solution known as a Reissner-Nordstr{\"o}m solution,
\begin{align}
  g_{\mu \nu}^{(0)}dx^\mu dx^\nu=-f_{\rm RN}dt^2 +f_{\rm RN}^{-1}dr^2+r^2d\Omega^2,\label{eq:RNmetric}
\end{align}
where $f_{\rm RN}:=f+{Q^2}/{r^2}$, and with
\begin{align}
    A_\mu =(-Q / r, 0,0,0).
\end{align}
Harmonic coordinates~$x^\mu$ on the Reissner-Nordstr{\"o}m spacetime are made by the collections of the scalar fields,
\begin{align}
    x^0:=&t,~x^1=\bar{r} \sin \theta \cos\varphi,\nonumber\\
    x^2:=&\bar{r} \sin \theta \sin\varphi,~x^3:=\bar{r} \cos\theta,~\bar{r}:=r-M,\label{eq:harmonic}
\end{align}
which satisfy $\Box x^\mu=0$.

\subsubsection{Perturbation equations in the low-frequency expansion}
We now consider linear perturbations, $g_{\mu \nu}=g_{\mu \nu}^{\rm (0)}+h_{\mu \nu}$, $A_\mu=A_{\mu}^{(0)} + \delta A_\mu $, within the low-frequency expansion in Appendix~\ref{App:GRsolution}. The metric perturbations are decomposed into the even and odd sectors of the form of Eqs.~\eqref{eq:Heven} and~\eqref{eq:hodd} but $f$ is replaced with $f_{\rm RN}$; the vector-field perturbations~$\delta A_\mu$ are decomposed into Eq.~\eqref{deltaA}. For the even-parity sector, the linear perturbations of the $i$th ($i=0,1$) order in the low-frequency expansion are governed by
\begin{align}
    {\cal L}_H\left[H^{(i)}\right]=&{\cal S}_H^{H,{\rm RN}}H^{(i)}+{\cal S}_H^{U,{\rm RN}}U^{(i)},\nonumber\\
    {\cal L}_{U} \left[U^{(i)}\right]=&{\cal S}_H^{U,{\rm RN}} U^{(i)}+{\cal S}_U^{H,{\rm RN}} H^{(i)},\label{eqsforHURN}
\end{align}
with
\begin{align}
 & {\cal S}_H^{H,{\rm RN}}:=\frac{2Q^2\left(r-M\right)}{r^4f f_{\rm RN}}\\
   &\times \left[\frac{d}{dr}-\frac{1}{rf}+\frac{1}{r}-\frac{\ell^2+\ell-2}{2\left(r-M\right)}-\frac{2\left(r-M\right)}{r^2 f_{\rm RN}}\right],\nonumber\\
  &{\cal S}_H^{U,{\rm RN}}:=\frac{4Q}{r^2 f_{\rm RN}}\left[\frac{d}{dr}+\frac{2\left(Q^2-Mr\right)}{r^3 f_{\rm RN}}\right],\\
&    {\cal S}_U^{U,{\rm RN}}:=-\frac{Q^2\left[\left(\ell^2+\ell+4\right)r-8M \right]}{r^5 f f_{\rm RN}},\\
&    {\cal S}_U^{H,{\rm RN}}:=\frac{Q}{r^2}\left[\frac{d}{dr}-\frac{2\left(Q^2-M r\right)}{r^3 f_{\rm RN}}\right].
\end{align}
Here, ${\cal L}_{H/U}$ are given in Eqs.~\eqref{eq:LHGR} and~\eqref{eq:LUGR}, respectively. 

For the odd-parity sector, the linear perturbation equations of the $i$th  order in the low-frequency expansion read
\begin{align}
    {\cal L}_h \left[h^{(i)}\right]=&{\cal S}_h^{h,{\rm RN}}h^{(i)}+{\cal S}_h^{u,{\rm RN}}u^{(i)},\nonumber\\
    ~{\cal L}_{u}\left[u^{(i)}\right]=&{\cal S}_u^{u,{\rm RN}}u^{(i)}+{\cal S}_u^{h,{\rm RN}}h^{(i)},\label{eqsforhuRN}
\end{align}
with
\begin{align}
   {\cal S}_h^{h,{\rm RN}}:=&-\frac{\left(\ell^2+\ell-2\right)Q^2}{r^4 f f_{\rm RN}},\\
    {\cal S}_h^{u,{\rm RN}}:=&-\frac{4Q}{r^2}\frac{d}{dr}u^{(i)},\\
   {\cal S}_u^{u,{\rm RN}}:=&\frac{2Q^2\left(r-M\right)}{r^4 f f_{\rm RN}}\left[\frac{d}{dr}-\frac{\ell\left(\ell+1\right)}{2\left(r-M\right)}\right],\\
   {\cal S}_u^{h,{\rm RN}}:=&-\frac{Q}{r^2 f_{\rm RN}}\left(\frac{d}{dr}-\frac{2}{r}\right).
\end{align}
The operators~${\cal L}_{h/u}$ are given in Eqs.~\eqref{eq:LhGR} and~\eqref{eq:LuGR}, respectively. 
The above equations are consistent with Eqs.~(35),~(36),~(39), and~(40) in Ref.~\cite{Cardoso:2017cfl}.\footnote{Equation~(36) in the literature contains a typo: the second term in ${\cal D}_{2}^{(2)}$ should be $(4Q^2-\eta r^2)/(r^4f)$. }

\subsubsection{General solution within perturbative expansions}
Now, we expand the perturbations in terms of $q(:=Q/M)$ to second order:
\begin{align}
    H^{(i)}\simeq & H^{(0,i)}+q^2 H^{(2,i)},\\
     h^{(i)}\simeq & h^{(0,i)}+q^2 h^{(2,i)},\\
     U^{(i)}\simeq & q U^{(1,i)},\\
         u^{(i)}\simeq & q u^{(1,i)}.
\end{align}
Equations~\eqref{eqsforHURN} and~\eqref{eqsforhuRN} reduce to Eq.~\eqref{eqforHh} at ${\cal O}(q^0)$. We then obtain
\begin{align}
    {\cal L}_U\left[ U^{(1,i)}\right]=& {\cal S}_U^{{\rm RN},(1,i)}H^{(0,i)},\nonumber\\
    ~{\cal L}_u\left[ u^{(1,i)}\right]=& {\cal S}_u^{{\rm RN},(1,i)}h^{(0,i)},
\end{align}
with
\begin{align}
    {\cal S}_U^{{\rm RN},(1,i)}:=&\frac{M}{r^2}\left(\frac{d}{dr}+\frac{2M}{r^2f}\right),\\
    {\cal S}_u^{{\rm RN},(1,i)}:=&-\frac{M}{r^2f}\left(\frac{d}{dr}-\frac{2}{r}\right),
\end{align}
at ${\cal O}(q)$, and 
\begin{align}
   {\cal L}_H \left[H^{(2,i)}\right]=&{\cal S}_H^{H,{\rm RN},(2,i)}H^{(0,i)}+{\cal S}_H^{U,{\rm RN},(1,i)}U^{(1,i)},\nonumber\\
   {\cal L}_h \left[h^{(2,i)}\right]=&{\cal S}_h^{h,{\rm RN},(2,i)}h^{(0,i)}+{\cal S}_h^{u,{\rm RN},(1,i)}u^{(1,i)},
\end{align}
with
\begin{align}
    &{\cal S}_H^{H,{\rm RN},(2,i)}:=\frac{2M^2\left(r-M\right)}{r^4 f^2}\left[\frac{d}{dr}-\frac{2+\ell\left(\ell+1\right)f}{2\left(r-M\right)f}\right],\\
     &{\cal S}_H^{U,{\rm RN},(1,i)}:=\frac{4M}{r^2f}\left(\frac{d}{dr}-\frac{2M}{r^2f}\right),\\
     & {\cal S}_h^{h,{\rm RN},(2,i)}:=-\frac{\left(\ell^2+\ell-2\right)M^2}{r^4f^2},\\
   & {\cal S}_h^{u,{\rm RN},(1,i)}:=-\frac{4M}{r^2}\frac{d}{dr},
\end{align}
at ${\cal O}(q^2)$. 

The general solutions for $H^{(0,i)}$ and $h^{(0,i)}$ take the form of Eq.~\eqref{eq:HhGR}. The general solutions for $U^{(1,i)}$ and $u^{(1,i)}$ are of the form
\begin{align}
    U^{(1,i)}=& \mathbb{E}_U^{(1,i)} U_T+ \mathbb{I}_U^{(1,i)}U_R+P_U^{(1,i)},\\
    u^{(1,i)}=&\mathbb{E}_u^{(1,i)} u_T+ \mathbb{I}_u^{(1,i)}u_R+P_u^{(1,i)},
\end{align}
where $\mathbb{E}_{U/u}^{(1,i)}$ and $\mathbb{I}_{U/u}^{(1,i)}$ are integration constants; $U_{T/R}$ and $u_{T/R}$ are defined in Eqs.~\eqref{eq:UT}--\eqref{eq:uR}; $P_{U/u}^{(1,i)}=P_{U/u}^{(1,i)}(r)$ are particular solutions. We do not write the explicit forms of $P_{U/u}^{(1,i)}$ as the expressions are cumbersome. The general solutions for $H^{(2,i)}$ and $h^{(2,i)}$ take the form of Eqs.~\eqref{eq:H2i} and~\eqref{eq:h2i}.

\subsubsection{Horizon-regular solution}
Imposing a regularity condition at the BH horizon~$r=M(1+\sqrt{1-q^2})$ specifies the linear perturbations. Now, we introduce the ingoing Eddington-Finkelstein coordinates~$(v,r,\theta,\varphi)$, where
\begin{align}
    v:=&t+r_{\rm RN}^*,\\
    r_{\rm RN}^*:=&r+2M\ln \left(\frac{r}{2M}-1\right)+\frac{q^2M^2}{r-2M}+{\cal O}\left(q^4\right).\nonumber
\end{align}

Let us denote $H, h, U, u$ as $X_\omega=X_\omega(r)$.  The Fourier transform of $X_\omega$ with respect to $t$ and that with respect to $v$ are related via
\begin{align}
    X_\omega e^{-i \omega t}\simeq X_\omega \left(1+i \omega r_{\rm RN}^*\right) e^{-i \omega v}, 
\end{align}
under the assumption of low frequencies. Eliminating logarithmically divergent terms in $X_\omega \left(1+i \omega r_{\rm RN}^*\right)$ around the horizon, we obtain horizon-regular solutions by determining the integration constants~$\mathbb{I}_{H/h}^{(i,j)}$ and $\mathbb{I}_{U/u}^{(1,j)}$ in terms of $\mathbb{E}_{H/h}^{(i,j)}$ and $\mathbb{E}_{U/u}^{(1,j)}$. When evaluating the CTRF~\eqref{eq:CTRFs}, we use Eqs.~\eqref{eq:E0HE1H} and~\eqref{eq:E0hE1h}. We also assume the absence of an external tidal electric field, i.e., $\mathbb{E}_{U/u}^{(1,j)}=0$, following Ref.~\cite{Cardoso:2017cfl}.

\subsection{Dynamical Chern-Simons gravity }
Dynamical Chern-Simons (dCS) gravity is an extension of GR, in which a pseudo-scalar field is coupled with a parity-violating quadratic curvature invariant (Pontryagin density) with a coupling constant $\alpha_\mathrm{dCS}$. The action is given by~\cite{Cardoso:2009pk,Alexander:2009tp,Molina:2010fb,Cardoso:2017cfl}
\begin{align}
    S_{\mathrm{dCS}}=\int d^{4} x \sqrt{-g}\left[R+\frac{\alpha_{\mathrm{dCS}}}{4} \Phi {}^*R_{\mu\nu\rho\lambda}R^{\nu\mu\rho\lambda}-\frac{1}{2} \partial_{\mu} \Phi \partial^{\mu} \Phi\right].
\end{align}
Note the following dimensions: $[\alpha_{\rm dCS}]=L^2$ and $[\Phi]=L^0$ where $L$ indicates the unit of length. Here, we have introduced
\begin{align}
    { }^{*} R^{\mu\nu\rho\lambda}:=\frac{1}{2} \epsilon^{\mu\nu\alpha\beta} R_{\alpha\beta}{}^{\rho\lambda}. \label{eq:RiemannStar}
\end{align}
The field equations read
\begin{align}
    R_{\mu \nu} & =\frac{1}{2} \partial_{\mu} \Phi \partial_{\nu} \Phi-\alpha_{\mathrm{dCS}} C_{\mu \nu}, \label{eq:gindCS} \\
    \square \Phi & =-\frac{\alpha_{\mathrm{dCS}}}{4} {}^*R_{\mu\nu\rho\lambda}R^{\nu\mu\rho\lambda}, \label{eq:PhiindCS}
\end{align}
with
\begin{align}
    C^{\mu \nu}:=(\nabla_{\rho} \Phi) \epsilon^{\rho \sigma \gamma(\mu} \nabla_{\gamma} R_{~\sigma}^{\nu)}+(\nabla_{\gamma} \nabla_{\rho} \Phi)^{*} R^{\gamma \mu  \nu \rho}.
\end{align}
The Schwarzschild spacetime is a solution to the above field equations as it satisfies ${}^*R_{\mu\nu\rho\lambda}R^{\mu\nu\rho\lambda}=0$ and $C_{\mu \nu}=0$. In such a background spacetime, Eqs.~\eqref{eq:gindCS} and~\eqref{eq:PhiindCS} reduce to the Einstein-massless Klein-Gordon system. In this case, the so-called no-hair theorem holds~\cite{PhysRevLett.28.452,Bekenstein:1995un}, meaning $\Phi=0$ on the background. Harmonic coordinates~$x^\mu$ on the Schwarzschild spacetime are made by the collections of the scalar fields in Eq.~\eqref{eq:harmonic}, which satisfy $\Box x^\mu=0$.  

\subsubsection{Perturbation equations in the low-frequency expansion}
We consider linear perturbations,~$g_{\mu\nu}=g_{\mu\nu}^{(0)}+h_{\mu \nu}$, $\Phi=0+\delta \Phi$, in the low-frequency expansion in Appendix~\ref{App:GRsolution}. For the even-parity sector, the perturbation equations are identical to those in vacuum GR~\cite{Cardoso:2009pk,Alexander:2009tp,Molina:2010fb,Cardoso:2017cfl}. For the odd-parity sector, the linear perturbations of the $i$th ($i=0,1$) order in the low-frequency expansion are governed by
\begin{align}
{\cal L}_h\left[h^{(i)}\right]=&\zeta_{\rm dCS}{\cal S}_{h}^{\rm dCS}\phi^{(i)},~{\cal L}_\phi\left[\phi^{(i)}\right]=\zeta_{\rm dCS} {\cal S}_\phi^{\rm dCS}h^{(i)},\label{eqforhphiindCS}
\end{align}
with
\begin{align}
    {\cal S}_h^{\rm dCS}:=&\frac{6 M^3 }{r^4}\left(\frac{d}{dr}-\frac{2}{r}\right),\\
    {\cal S}_\phi^{\rm dCS}:=&\frac{6\ell\left(\ell+1\right)M^3}{r^4f}\left(\frac{d}{dr}-\frac{2}{r}\right),
\end{align}
where $\zeta_{\rm dCS}:=\alpha_{\rm dCS}/M^2$. Note that $\phi^{(i)}$ is of ${\cal O}(\zeta_{\rm dCS})$. Here, ${\cal L}_h$ and ${\cal L}_\phi$ are defined in Eqs.~\eqref{eq:LhGR} and~\eqref{eq:LphiGR}, respectively. The identical equations to Eq.~\eqref{eqforhphiindCS} can also be derived with the Mathematica notebook provided in Ref.~\cite{Pani:2013pma}.

\subsubsection{General solution within perturbative expansions}
Let us expand $h^{(i)}$ and $\phi^{(i)}$ in $\zeta_{\rm dCS}$ to ${\cal O}(\zeta_{\rm dCS}^2)$:
\begin{align}
    h^{(i)}\simeq & h^{(0,i)}+\zeta_{\rm dCS}^2 h^{(2,i)},\\
    \phi^{(i)}\simeq & \zeta_{\rm dCS} \phi^{(1,i)}.
\end{align}
Then, the linear perturbations in Eq.~\eqref{eqforhphiindCS} reduce to the form in Eq.~\eqref{eqforHh} at ${\cal O}(\zeta_{\rm dCS}^0)$, and 
\begin{align}
{\cal L}_\phi\left[\phi^{(1,i)}\right]=  {\cal S}_\phi^{\rm dCS}h^{(0,i)},
\end{align}
at ${\cal O}(\zeta_{\rm dCS})$, and
\begin{align}
{\cal L}_h\left[h^{(2,i)}\right]=& {\cal S}_h^{\rm dCS}\phi^{(1,i)},\label{eq:h2iindCS}
\end{align}
at ${\cal O}(\zeta_{\rm dCS}^2)$.

The general solution for $h^{(0,i)}$ is given in the form of Eq.~\eqref{eq:HhGR} at ${\cal O}(\zeta_{\rm dCS}^0)$. The general solution for $\phi^{(1,i)}$ takes the form
\begin{align}
    \phi^{(1,i)}=& \mathbb{E}_\phi^{(1,i)} \phi_T+ \mathbb{I}_\phi^{(1,i)}\phi_R+P_\phi^{(1,i)},
\end{align}
where $\mathbb{E}_\phi^{(1,i)}$ and $\mathbb{I}_\phi^{(1,i)}$ are integration constants; $\phi_{T/R}$ are defined in Eqs.~\eqref{eq:phiT} and~\eqref{eq:phiR}; $P_{\phi}^{(1,i)}=P_{\phi}^{(1,i)}(r)$ is a particular solution. The general solution for $h^{(2,i)}$ takes the form in Eq.~\eqref{eq:h2i}. We assume the absence of an external tidal scalar field, $\mathbb{E}_{\phi}^{(1,i)}=0$, following Ref.~\cite{Cardoso:2017cfl}. To solve Eq.~\eqref{eq:h2iindCS} in closed form, we expand $\phi^{(1,i)}$ in powers of $M/r$ to ${\cal O}((M/r)^{10})$. We have checked that the relative deviation of the results is of ${\cal O}(10^{-1})\%$ when truncating at ${\cal O}((M/r)^{9})$ or ${\cal O}((M/r)^{11})$.  This approximation allows us to estimate the TLNs and TDNs presented in Tables~\ref{table:TLNs} and~\ref{table:TDNs}.

\subsubsection{Horizon-regular solution}
Imposing a regularity condition at the BH horizon~$r=2M$ specifies the linear perturbation with the ingoing Eddington-Finkelstein coordinates~$(v,r,\theta,\varphi)$, where
\begin{align}
    v:=&t+r^*,\\
    r^*:=&r+2M\ln \left(\frac{r}{2M}-1\right).\nonumber
\end{align}
We denote $h,\phi$ as $X_\omega=X_\omega(r)$. By eliminating logarithmically divergent terms in $X_\omega \left(1+i \omega r^*\right)$ around the horizon, we obtain horizon-regular solutions. We have confirmed that the horizon-regular solution for $\phi^{(1,0)}$ agrees with Eq.~(52) in Ref.~\cite{Cardoso:2017cfl}. Thus, we determine~$\mathbb{I}_{h}^{(i,j)}$ and $\mathbb{I}_{\phi}^{(1,j)}$ in terms of $\mathbb{E}_h^{(i,j)}$ and $\mathbb{E}_\phi^{(1,j)}$. When computing the CTRF~\eqref{eq:CTRFs}, we use Eqs.~\eqref{eq:E0HE1H} and~\eqref{eq:E0hE1h}.

\subsection{Einstein-dilaton-Gauss-Bonnet gravity}
A similar (but another) extension of GR is the Einstein-dilaton-Gauss-Bonnet gravity, in which a scalar field is now coupled with an even-parity quadratic curvature invariant with a coupling constant $\alpha_\mathrm{dGB}$. The action is given by~\cite{Kanti:1995vq,Blazquez-Salcedo:2016enn,Bryant:2021xdh} 
\begin{align}
    S_{\mathrm{dGB}}=\frac{1}{16 \pi} \int \mathrm{d}^{4} x \sqrt{-g}\left[R-\frac{1}{2} \partial_{\mu} \Phi \partial^{\mu} \Phi+\alpha_{\rm dGB} e^\Phi \mathscr{G}  \right],
\end{align}
with the Gauss-Bonnet term given by
\begin{align}
    \mathscr{G}:=R_{\mu \nu \rho\sigma} R^{\mu \nu \rho \sigma}-4 R_{\mu \nu} R^{\mu \nu}+R^{2}.
\end{align}
The field equations read\footnote{The right-hand side of Eq.~(3) in Ref.~\cite{Blazquez-Salcedo:2016enn} and in Eq.~(2.3) in Ref.~\cite{Bryant:2021xdh} is missing a minus sign. These are typos and we have verified that the actual computations in the literature are correct.}
\begin{align}
    G_{\mu \nu}=& \frac{1}{2} \partial_{\mu} \Phi \partial_\nu \Phi-\frac{1}{4} g_{\mu \nu} \left(\partial_\rho \Phi \right)\left(\partial^\rho \Phi\right)-\alpha_{\rm dGB} \mathcal{K}_{\mu \nu} ,\label{EinstenwithdGB}\\
    \Box \Phi=& -\alpha_{\rm dGB} e^\Phi \mathscr{G} ,\label{dilaton} 
\end{align}
with
\begin{align}
    \mathcal{K}_{\mu \nu}:=2\left(g_{\mu \rho} g_{\nu \sigma}+g_{\mu \sigma} g_{\nu \rho}\right) \epsilon^{\alpha \sigma \beta \delta} \nabla_{\gamma}\left[{ }^{*} R^{\rho \gamma}{ }_{\beta \delta} \partial_{\alpha} e^{\Phi}\right].
\end{align}
Here ${ }^{*} R^{\mu \nu}{ }_{\rho\sigma}$ is the dual of the Riemann tensor in Eq.~\eqref{eq:RiemannStar}.

We expand Eqs.~\eqref{EinstenwithdGB} and~\eqref{dilaton} in terms of $\zeta_{\rm dGB}:=\alpha_{\rm dGB}/M^2$ up to second order perturbatively. We thus find a static and spherically symmetric solution,
\begin{align}
    g_{\mu\nu}^{(0)} dx^\mu dx^\nu=-f_t^{\rm dGB}dt^2+\frac{1}{f_r^{\rm dGB}}dr^2+r^2d\Omega^2,
\end{align}
where
\begin{widetext}
\begin{align}
    f_t^{\rm dGB}:=&f-\zeta_{\rm dGB}^2\frac{M}{r}\left(\frac{49}{40}-\frac{M^2}{3r^2}-\frac{26M^3}{3r^3}-\frac{22M^4}{5r^4}-\frac{32M^5}{5r^5}+\frac{80M^6}{3r^6}\right),\\
    f_r^{\rm dGB}:=&f-\zeta_{\rm dGB}^2\frac{M}{r}\left(\frac{49}{40}-\frac{M}{r}-\frac{M^2}{r^2}-\frac{52M^3}{3r^3}-\frac{2M^4}{r^4}-\frac{16M^5}{5r^5}+\frac{368M^6}{3r^6}\right),
\end{align}
with a dilaton field,
\begin{align}
\Phi^{(0)}=\frac{2M}{r}\zeta_{\rm dGB} \left(1+\frac{M}{r}+\frac{4M^2}{3r^2}\right) +\zeta_{\rm dGB}^2\left(\frac{73M}{30r}+\frac{73M^2}{30r^2}+\frac{146M^3}{45r^3}+\frac{73M^4}{15r^4}+\frac{224M^5}{75r^5}+\frac{16M^6}{9r^6}\right).
\end{align}
\end{widetext}
Note that the Arnowitt--Deser--Misner~(ADM) mass of this perturbative spacetime receives a correction to the Schwarzschild ADM mass~$M$, i.e., $M(1+\zeta_{\rm dGB}^2 49/80)$~\cite{Bryant:2021xdh}.
The BH horizon location is $r=2M$. These solutions are in perfect agreement with findings in Ref.~\cite{Bryant:2021xdh}. Harmonic coordinates~$x^\mu$ on the above spacetime are made by the collections of the scalar fields,
\begin{align}
    x^0:=&t,~x^1=\bar{r} \sin \theta \cos\varphi,\\
    x^2:=&\bar{r} \sin \theta \sin\varphi,~x^3:=\bar{r} \cos\theta,~\bar{r}:=r-M+\zeta_{\rm dGB}^2 \delta \bar{r}_{\rm dGB},\nonumber
\end{align}
which satisfy $\Box x^\mu={\cal O}(\zeta_{\rm dGB}^3)$. The explicit form of $\delta \bar{r}_{\rm dGB}$ is not relevant for later discussion.

\subsubsection{Perturbation equations in the low-frequency expansion}
When considering linear perturbations, $g_{\mu \nu}=g_{\mu \nu}^{(0)}+h_{\mu \nu}$, $\Phi=\Phi^{(0)}+\delta \Phi$, within the low-frequency expansion in the manner of Appendix~\ref{App:GRsolution}, the even-parity sector of the linear perturbations of the $i$th ($i=0,1$) order in the low-frequency expansion is governed by
\begin{align}
    &{\cal L}_H\left[ H^{(i)}\right]=\zeta_{\rm dGB} {\cal S}_H^{{\rm dGB},(1,i)} \phi^{(i)}+\zeta_{\rm dGB}^2 {\cal S}_H^{H,{\rm dGB},(2,i)} H^{(i)},\label{eqforHindEdGB}\\
    &{\cal L}_\phi\left[ \phi^{(i)}\right]=\zeta_{\rm dGB} {\cal S}_\phi^{H,{\rm dGB}} H^{(i)},\label{eqforphiindEdGB}
\end{align}
where
\begin{widetext}
\begin{align}
    &{\cal S}_H^{{\rm dGB},(1,i)}:= \frac{48M^3 \left(r-M\right)}{r^6f}\left[\frac{d}{dr}+\frac{r^3-6\left(\sigma_\ell+2\right)Mr^2+12\left(\sigma_\ell+1\right)M^2 r+8M^3}{12Mr \left(r^2-3Mr+2M^2\right)}   \right],\\
     &{\cal S}_H^{H,{\rm dGB},(2,i)}:=-\frac{M\left(147r^6+174Mr^5-5532M^2r^4+9896M^3r^3-1568M^4r^2+180800M^5r-453120M^6\right)}{120r^8 f}\\
    & \times \left[\frac{d}{dr}-\frac{c_H^{{\rm dGB},(7)}r^7 +c_H^{{\rm dGB},(6)}r^6+c_H^{{\rm dGB},(5)}r^5+c_H^{{\rm dGB},(4)}r^4+c_H^{{\rm dGB},(3)}r^3+c_H^{{\rm dGB},(2)}r^2+c_H^{{\rm dGB},(1)}r +506880M^7 }{r^2 f \left(147r^6+174Mr^5-5532M^2r^4+9896M^3r^3-1568M^4r^2+180800M^5r-453120M^6\right)}\right],\nonumber\\
    &{\cal S}_\phi^{H,{\rm dGB}}:=-\frac{48M^3\left(r-3M\right)r^2}{r^6f} \left[\frac{d}{dr}+\frac{r^3-6\sigma_\ell Mr^2+12\left(\sigma_\ell+1\right)M^2r-56M^3}{12M\left(r-3M\right)r^2f}\right],
\end{align}
\end{widetext}
with
\begin{align}
    c_H^{{\rm dGB},(7)}=& 147 \sigma_\ell  ,\nonumber\\ 
     c_H^{{\rm dGB},(6)}=& -12\left(10\sigma_\ell-49\right)M  ,\nonumber\\ 
      c_H^{{\rm dGB},(5)}=& -24\left(125\sigma_\ell -49 \right)M^2  ,\nonumber\\ 
       c_H^{{\rm dGB},(4)}=&  16\left(170\sigma_\ell +117\right)M^3 ,\\ 
        c_H^{{\rm dGB},(3)}=& -16 \left(15\sigma_\ell +2446\right) M^4 ,\nonumber\\ 
         c_H^{{\rm dGB},(2)}=&128\left(717\sigma_\ell -199\right)M^5   ,\nonumber\\ 
          c_H^{{\rm dGB},(1)}=& -640 \left(253\sigma_\ell +10\right)M^6.\nonumber
\end{align}
Here, we have defined $\sigma_\ell:=\ell(\ell+1)$. Note that $\phi^{(i)}$ is of ${\cal O}(\zeta_{\rm dGB})$, and hence, the lowest order of the right-hand side of Eq.~\eqref{eqforHindEdGB} is second order. The operators~${\cal L}_H$ and ${\cal L}_\phi$ are given in Eqs.~\eqref{eq:LHGR} and~\eqref{eq:LphiGR}, respectively. 

In the odd-parity sector, the dilaton field is not perturbed and the metric perturbation of the $i$th order in the low-frequency expansion is described by
\begin{align}
    {\cal L}_h\left[h^{(i)}\right]=\zeta_{\rm dGB}^2{\cal S}_h^{\rm dGB}h^{(i)},\label{eqforhindEdGB}
\end{align}
with
\begin{widetext}
\begin{align}
    &{\cal S}_h^{\rm dGB}:= \left(\frac{M^2}{r^3}+\frac{28M^3}{r^4}+\frac{28M^4}{r^5}+\frac{64M^5}{r^6}-\frac{240M^6}{r^7} \right)\nonumber\\
&\times \left[\frac{d}{dr}+\frac{c_h^{{\rm dGB},(5)}r^5+c_h^{{\rm dGB},(4)}r^4 + c_h^{{\rm dGB},(3)}r^3+  c_h^{{\rm dGB},(2)}r^2 + c_h^{{\rm dGB},(1)} r+57600M^5}{120M r \left(r^4+28M r^3+28M^2r^2+64M^3 r-240M^4\right)}\right],    
\end{align}
    \end{widetext}
where
\begin{align}
    c_h^{{\rm dGB},(5)}=&147\left(\sigma_\ell-2\right),\nonumber\\
    c_h^{{\rm dGB},(4)}=&12\left(39\sigma_\ell-98\right)M,\nonumber\\
    c_h^{{\rm dGB},(3)}=&-12\left(143\sigma_\ell+274\right)M^2,\\
    c_h^{{\rm dGB},(2)}=&-64\left(94\sigma_\ell-83\right)M^3,\nonumber\\
    c_h^{{\rm dGB},(1)}=&-160\left(109\sigma_\ell-122\right)M^4.\nonumber
\end{align}

\subsubsection{General solution within perturbative expansions}
We expand the perturbations in $\zeta_{\rm dGB}$ to ${\cal O}(\zeta_{\rm dGB}^2)$:
\begin{align}
    H^{(i)}\simeq& H^{(0,i)}+\zeta_{\rm dGB}^2 H^{(2,i)},\\
    h^{(i)}\simeq& h^{(0,i)}+\zeta_{\rm dGB}^2 h^{(2,i)},\\
    \phi^{(i)}\simeq& \zeta_{\rm dGB} \phi^{(1,i)}.
\end{align}
Equations~\eqref{eqforHindEdGB} and~\eqref{eqforhindEdGB}  reduce to Eqs.~\eqref{eqforHh} and~\eqref{eqforphi} at ${\cal O}(\zeta_{\rm dGB}^0)$, and 
\begin{align}
    {\cal L}_\phi\left[ \phi^{(1,i)}\right]=&{\cal S}_\phi^{H,{\rm dGB}} H^{(0,i)},
\end{align}
at ${\cal O}(\zeta_{\rm dGB})$,
and
\begin{align}
    {\cal L}_H\left[ H^{(2,i)}\right]=&{\cal S}_H^{{\rm dGB},(1,i)} \phi^{(1,i)}+{\cal S}_H^{H,{\rm dGB},(2,i)} H^{(0,i)},\label{eq:eqforH2inEdGB}\\
     {\cal L}_h\left[ h^{(2,i)}\right]=& {\cal S}_h^{h,{\rm dGB}} h^{(0,i)},
\end{align}
at ${\cal O}(\zeta_{\rm dGB}^2)$.

The general solutions for $H^{(0,i)}$ and $h^{(0,i)}$ are in the form of Eq.~\eqref{eq:HhGR}. The general solution for $\phi^{(1,i)}$ takes the form
\begin{align}
    \phi^{(1,i)}=& \mathbb{E}_\phi^{(1,i)} \phi_T+ \mathbb{I}_\phi^{(1,i)}\phi_R+P_\phi^{(1,i)},
\end{align}
where $\mathbb{E}_\phi^{(1,i)}$ and $\mathbb{I}_\phi^{(1,i)}$ are integration constants; $\phi_{T/R}$ are defined in Eqs.~\eqref{eq:phiT} and~\eqref{eq:phiR}; $P_{\phi}^{(1,i)}=P_{\phi}^{(1,i)}(r)$ is a particular solution. The general solutions for $H^{(2,i)}$ and $h^{(2,i)}$ take the form of Eq.~\eqref{eq:h2i}. To solve Eq.~\eqref{eq:eqforH2inEdGB} in closed form, we expand $H^{(0,1)}$ and $\phi^{(1,i)}$ in powers of $M/r$ to ${\cal O}((M/r)^{15})$, similar to the dCS case. Within this approximation, the results presented in Tables~\ref{table:TLNs} and~\ref{table:TDNs} are estimated. We have checked that the relative deviation of the results is of ${\cal O}(10^{-1})\%$ and ${\cal O}(10^{0})\%$ for the TLNs and the TDNs, respectively, when truncating at ${\cal O}((M/r)^{14})$ or ${\cal O}((M/r)^{16})$.

\subsubsection{Horizon-regular solution}
Imposing a regularity condition at the BH horizon~$r=2M$  specifies the linear perturbation with the ingoing Eddington-Finkelstein coordinates~$(v,r,\theta,\varphi)$, where
\begin{align}
    v:=&t+r^*,\\
    r^*:=&r+2M\ln \left(\frac{r}{2M}-1\right).\nonumber
\end{align}
Eliminating logarithmically divergent terms in $X_\omega \left(1+i \omega r^*\right)$ around the horizon with $X_\omega$ representing a collection of $(H,h,\phi)$, we obtain horizon-regular solutions, thereby determining the integration constants~$\mathbb{I}_{H/h}^{(i,j)}$ and $\mathbb{I}_{\phi}^{(1,j)}$ in terms of $\mathbb{E}_{H/h}^{(i,j)}$ and $\mathbb{E}_\phi^{(1,j)}$. In the computation of the CTRF~\eqref{eq:CTRFs}, we use Eqs.~\eqref{eq:E0HE1H} and~\eqref{eq:E0hE1h}. For simplicity, we assume that an external scalar tidal field is absent,~$\mathbb{E}_\phi^{(1,i)}=0$.

\subsection{Braneworld}
The braneworld scenario is an extension of GR that incorporates an extra spatial dimension. A vacuum solution in the theory is described by a solution within GR in four dimensions with a trace-free energy-momentum tensor that satisfies~\cite{Dadhich:2000am,Maartens:2003tw}
\begin{align}
G_{\mu \nu}=E_{\mu \nu}.
\end{align}
Here, $E_{\mu \nu}$ is a tensor field associated with the bulk Weyl tensor and takes the form~(on a spherically symmetric background on the brane),
\begin{align}
    E_{\mu \nu}=\rho \left(\frac{1}{3}g_{\mu\nu} +\frac{2}{3}U_{\mu}U_{\nu}\right)+\Pi\left(\frac{1}{3}g_{\mu \nu}+\frac{1}{3}U_\mu U_\nu -r_\mu r_\nu\right).
\end{align}
The vectors~$U_\mu$ and $r_\mu$ are unit timelike and radial vectors, respectively, and $\rho=\rho(r)$, $\Pi=\Pi(r)$. One finds a static and spherically symmetric solution of the same form as the Reissner-Nordstr{\"o}m metric~\eqref{eq:RNmetric}~\cite{Maartens:2003tw},
\begin{align}
    g_{\mu \nu}^{(0)} dx^\mu dx^\nu =-f_{\rm B}dt^2+f^{-1}_{\rm B}dr^2+r^2d\Omega^2,
\end{align}
where $f_{\rm B}:= f+\beta/r^2$ with 
\begin{align}
    \rho=\frac{\beta}{8\pi r^4},~\Pi=2\rho.
\end{align}
Here, $\beta$ is a constant called a tidal charge. Harmonic coordinates~$x^\mu$ on this background are made by the collections of the scalar fields in Eq.~\eqref{eq:harmonic}, which satisfy $\Box x^\mu=0$.

\subsubsection{Perturbation equations in the low-frequency expansion}
Consider linear perturbations, $g_{\mu \nu}=g_{\mu \nu}^{(0)} +h_{\mu \nu}$, $\rho\to \rho +\delta \rho(r)Y_{\ell m}e^{-i\omega t}$, $\Pi\to \Pi+\delta \Pi(r) Y_{\ell m}e^{-i\omega t}$, and even for the vectors, $U_\mu \to U_\mu+\delta U_\mu$ and $r_\mu \to r_\mu+\delta r_\mu$, where their harmonic decompositions are
\begin{align}
\label{deltaU}
    \left(\delta U_{\ell m}\right)_\mu =&\begin{pmatrix}
    \delta U_t^{\ell m} Y_{\ell m} e^{-i\omega t}\\
   \delta U_r^{\ell m} Y_{\ell m} e^{-i\omega t}\\
     U_k^{\ell m} \partial_\theta Y_{\ell m} e^{-i\omega t}+\dfrac{U_u^{\ell m} }{\sin\theta} \partial_\varphi Y_{\ell m}  e^{-i\omega t}\\
      U_k^{\ell m} \partial_\varphi Y_{\ell m} -U_u^{\ell m} \sin\theta \partial_\theta Y_{\ell m} e^{-i\omega t}
     \end{pmatrix},\\
       \left(\delta r_{\ell m}\right)_\mu =&\begin{pmatrix}
    \delta r_t^{\ell m} Y_{\ell m} e^{-i\omega t}\\
   \delta r_r^{\ell m} Y_{\ell m} e^{-i\omega t}\\
     r_k^{\ell m} \partial_\theta Y_{\ell m} e^{-i\omega t}+\dfrac{r_u^{\ell m} }{\sin\theta} \partial_\varphi Y_{\ell m}  e^{-i\omega t}\\
      r_k^{\ell m} \partial_\varphi Y_{\ell m} -r_u^{\ell m} \sin\theta \partial_\theta Y_{\ell m} e^{-i\omega t}
     \end{pmatrix}. 
\end{align}
The normalization conditions~$(U+\delta U)_\mu (U+\delta U)^\mu=-1$ and $(r+\delta r)_\mu (r+\delta r)^\mu=1$ within linear order of the perturbation allow us to express $\delta U_t^{\ell m}$ and $\delta r_r^{\ell m}$ in terms of the metric perturbations. We assume $\delta r_t=0$, allowing us to express $\delta U_r^{\ell m}$ in terms of the metric and matter-density perturbations. In the low-frequency expansion in Appendix~\ref{App:GRsolution}, the even-parity sector of the perturbations of the $i$th ($i=0,1$) order in the low-frequency expansion is governed by
\begin{align}
    {\cal L}_H \left[H^{(i)}\right]={\cal S}_H^{H,{\rm B}}H^{(i)}+{\cal S}_H^{\delta\rho,{\rm B}}\delta \rho^{(i)},\label{eqsforHB}
\end{align}
and \begin{align}
    \left[\frac{d}{dr}+\frac{2\left(r-M\right)}{\beta+r^2 f} \right]\delta \rho^{(i)}=\frac{\beta}{4\pi r^2}\left[\frac{d}{dr}-\frac{2\left(\beta-Mr\right)}{r^3 \left(\beta+r^2 f \right)}\right]H^{(i)}.\label{eqsfordeltarho}
\end{align}
Here, ${\cal L}_H$ is defined in Eq.~\eqref{eq:LHGR}  and we have introduced
\begin{align}
    &{\cal S}_H^{H,{\rm B}}:=-\frac{2\beta \left[r^3 -5M r^2+\left(5M^2+2\beta\right)r-3M\beta\right]}{r^4 f f_{\rm B} \left(Mr-\beta\right)}\left(\frac{d}{dr}-\Sigma^{\rm B} \right),\nonumber\\
       &{\cal S}_H^{\delta\rho,{\rm B}}:=\frac{8\pi r^3}{Mr-\beta}\left(\frac{d}{dr}+\frac{2}{r}\right),
\end{align}
where 
\begin{align}
    \Sigma^{\rm B}:=&\frac{1}{rf}+\frac{1}{r}-\frac{8r^2+\left(\sigma_\ell-26\right)Mr-\left(\sigma_\ell-6\right)\beta +12 M^2}{2 \left[r^3 -5M r^2+\left(5M^2+2\beta\right)r-3M\beta\right]}\\
    &+\frac{2\left(r-M\right)}{r^2 f_{\rm B}}. \nonumber
    \end{align}
    
For the odd-parity sector, the perturbation of the $i$th order in the low-frequency expansion is described by
\begin{align}
    {\cal L}_h\left[h^{(i)}\right]={\cal S}_h^{\rm B}h^{(i)},\label{eqsforhB}
\end{align}
with
\begin{align}
    {\cal S}_h^{\rm B}:=-\beta\frac{\sigma_\ell-2}{r^2 f_{\rm B}\left(\beta +r^2 f_{\rm B}\right)}.
\end{align}
The operator~${\cal L}_h$ is given by Eq.~\eqref{eq:LhGR}.

\subsubsection{General solution within perturbative expansions}
We now introduce $\zeta_{\rm B}:=\beta/M^2$ and expand the linear perturbations to ${\cal O}(\zeta_{\rm B})$:
\begin{align}
H^{(i)}\simeq &H^{(0,i)}+\zeta_{\rm B}H^{(1,i)},\\
h^{(i)}\simeq &h^{(0,i)}+\zeta_{\rm B}h^{(1,i)},\\
\delta\rho^{(i)}\simeq &\zeta_{\rm B}\delta\rho^{(1,i)}.
\end{align}
Equations~\eqref{eqsforHB} and~\eqref{eqsforhB} reduce to Eq.~\eqref{eqforHh} at ${\cal O}(\zeta_{\rm B}^0)$. We then obtain
\begin{align}&\left[\frac{d}{dr}+\frac{2\left(r-M\right)}{r^2 f}\right]\delta \rho^{(1,i)}=\frac{M^2}{4\pi r^4}\left(\frac{d}{dr}+\frac{2M}{r^2 f}\right)H^{(0,i)},\label{eqsfordeltarho1}\\
    &{\cal L}_H \left[H^{(1,i)}\right]={\cal S}_H^{H,{\rm B},(1,i)}H^{(0,i)}+{\cal S}_H^{\delta\rho,{\rm B},(0,i)}\delta\rho^{(1,i)},\label{eqsforHB1}\\
    &{\cal L}_h\left[h^{(1,i)}\right]={\cal S}_h^{{\rm B},(1,i)}h^{(0,i)},\label{eqsforhB1}
\end{align}
at ${\cal O}(\zeta_{\rm B})$. Here, we have defined
\begin{align}
    &{\cal S}_H^{H,{\rm B},(1,i)}:=-\frac{2M \left(r^2-5Mr+5M^2\right)}{r^4f^2}\left(\frac{d}{dr}+\Sigma^{{\rm B},(1,i)}\right),\nonumber\\
   &{\cal S}_H^{\delta\rho,{\rm B},(0,i)}:=\frac{8\pi r^2}{M}\left(\frac{d}{dr}+\frac{2}{r}\right),\nonumber
   \end{align}
   with
   \begin{align}
       \Sigma^{{\rm B},(1,i)}:= \frac{M\left[\left(\sigma_\ell+6\right) r^2-2\left(\sigma_\ell+8\right)Mr+16M^2\right]}{2r^2 f \left(r^2-5M r +5M^2\right)},
   \end{align} 
   and
   \begin{align}
    &  {\cal S}_h^{{\rm B},(1.i)}:=-\frac{\left(\sigma_\ell-2\right)M^2}{r^4 f^2}.
\end{align}

The general solutions for $H^{(0,i)}$ and $h^{(0,i)}$ are given by Eq.~\eqref{eq:HhGR}. With the general solution for $\delta \rho^{(1,i)}$ at hand, the general solutions for $H^{(1,i)}$ and $h^{(1,i)}$ are derived in the form of Eqs.~\eqref{eq:H1i} and~\eqref{eq:h1i}. When solving Eq.~\eqref{eqsforHB1} at $(1,1)$th order, we expand $\delta \rho^{(1,1)}$ in powers of $M/r$ to ${\cal O}((M/r)^{35})$. Within this approximation, the results in Table~\ref{table:TDNs} for the even-parity sector are estimated. We have checked that, if the series is truncated at ${\cal O}((M/r)^{34})$ or ${\cal O}((M/r)^{36})$, then the relative deviation of the resulting TDNs is of ${\cal O}(10^{-1})\%$.

\subsubsection{Horizon-regular solution}
Imposing a regularity condition at the BH horizon~$r=M(1+\sqrt{1-\zeta_{\rm B}})$ specifies the linear perturbations. Now, we introduce the ingoing Eddington-Finkelstein coordinates~$(v,r,\theta,\varphi)$, where
\begin{align}
    v:=&t+r_{\rm B}^*,\\
    r_{\rm B}^*:=&r+2M\ln \left(\frac{r}{2M}-1\right)+\frac{\zeta_{\rm B}M^2}{r-2M}+{\cal O}\left(\zeta_{\rm B}^2\right).\nonumber
\end{align}
For $X_\omega=H, h, \delta\rho$, we obtain horizon-regular solutions by eliminating logarithmically divergent terms in $X_\omega \left(1+i \omega r_{\rm B}^*\right)$ around the horizon, which further determines the integration constants~$\mathbb{E}_{H/h}^{(i,j)}$. When evaluating the CTRF~\eqref{eq:CTRFs}, we use Eqs.~\eqref{eq:E0HE1H} and~\eqref{eq:E0hE1h}.

\subsection{EFT}
An effective field theory~(EFT) approach is the most general extension of GR under several assumptions~(see the details in Refs.~\cite{Endlich:2017tqa,Cardoso:2018ptl}). The action is given by\footnote{For simplicity, we here do not consider the $1/\Lambda_-^6$ term in the literature.}
\begin{align}
    S_{\mathrm{eff}} & =2\int d^{4} x \sqrt{-g} \left(R-\frac{\mathcal{C}^{2}}{\Lambda^{6}}-\frac{\tilde{\mathcal{C}}^{2}}{\tilde{\Lambda}^{6}}
    \right), \\
    \mathcal{C} & \equiv R_{\alpha \beta \gamma \delta} R^{\alpha \beta \gamma \delta}, \quad \tilde{\mathcal{C}} \equiv 2R_{\alpha \beta \gamma \delta} {}^*R^{\alpha \beta \gamma \delta},\nonumber
\end{align}
where ${}^* R^{\alpha \beta \gamma \delta}$ is the dual of the Riemann tensor~\eqref{eq:RiemannStar}. The parameters~$\Lambda$ and $\tilde{\Lambda}$ are coupling constants with the dimension of the inverse length. The field equations read
\begin{align}
    & G^{\mu \alpha}=\frac{1}{\Lambda^{6}}\left(8 R^{\mu \nu \alpha \beta} \nabla_{\nu} \nabla_{\beta} \mathcal{C}+\frac{g^{\mu \alpha}}{2} \mathcal{C}^{2}\right) \notag \\
    + & \frac{1}{\tilde{\Lambda}^{6}}\left(8 \tilde{R}^{\mu \rho \alpha \nu} \nabla_{\rho} \nabla_{\nu} \tilde{\mathcal{C}}+\frac{1}{2} g^{\mu \alpha} \tilde{\mathcal{C}}^{2}\right). 
\end{align}
Following Ref.~\cite{Cardoso:2018ptl}, we introduce dimensionless coupling constants normalized by the ADM mass of spacetime~$M$,
\begin{align}
    \left(\epsilon_1,\epsilon_2\right)=\left(\frac{1}{M^6 \Lambda^6}, \frac{1}{M^6 \tilde{\Lambda}^6}\right).
\end{align}

The background spacetime deviates from the Schwarzschild spacetime. We recover the solution derived by Ref.~\cite{Cardoso:2018ptl}:
\begin{align}
    g_{\mu \nu}^{(0)} dx^\mu dx^\nu=-f_t^{\rm EFT}dt^2+\frac{1}{f_r^{\rm EFT}}dr^2+r^2d\Omega^2,
\end{align}
where 
\begin{align}
    f_t^{\rm EFT}:=& f+\epsilon_1  \left(-\frac{1024M^9}{r^9}+\frac{1408M^{10}}{r^{10}}\right),\nonumber\\
    f_r^{\rm EFT}:=& f+\epsilon_1 \left(-\frac{4608M^9}{r^9}+\frac{8576M^{10}}{r^{10}}\right).
\end{align}
The event horizon is located at $r=2M(1+5\epsilon_1/8)+{\cal O}(\epsilon_1^2)$~\cite{Cardoso:2018ptl}. Harmonic coordinates~$x^\mu$ on the above spacetime are made by the collections of the scalar fields,
\begin{align}
    x^0:=&t,~x^1=\bar{r} \sin \theta \cos\varphi,\\
    x^2:=&\bar{r} \sin \theta \sin\varphi,~x^3:=\bar{r} \cos\theta,~\bar{r}:=r-M+\epsilon_1 \delta \bar{r}_{\rm EFT},\nonumber
\end{align}
which satisfy $\Box x^\mu={\cal O}(\epsilon_1^2)$. The explicit form of $\delta \bar{r}_{\rm EFT}$ is not relevant for later discussion.
 
\subsubsection{Perturbation equations in the low-frequency expansion}

In the metric perturbation in the manner of Appendix~\ref{App:GRsolution}, the even-parity sector of the $i$th ($i=0,1$) order in the low-frequency expansion is described by
\begin{align}
    {\cal L}_H\left[H^{(i)}\right]=\epsilon_1 {\cal S}_H^{\epsilon_1} H^{(i)},\label{eqforHinEFT} 
\end{align}
with
\begin{widetext}
\begin{align}
    &{\cal S}_H^{\epsilon_1}:=- 256M^8\frac{48\left(\sigma_\ell+9\right) r^3 -6\left(32\sigma_\ell+519\right)M r^2 +\left(192\sigma_\ell+7237\right)M^2 r-5481M^3}{r^{12} f^2} \left[\frac{d}{dr}\right.\\
    &\left.-\frac{3\sigma_\ell \left(3\sigma_\ell+170\right)r^4 -6\left(\sigma_\ell \left(6\sigma_\ell+533\right)+72\right)Mr^3 +  \left(\sigma_\ell \left(36\sigma_\ell +6689\right)+5454\right)M^2 r^2 -2\left(2333\sigma_\ell +8096\right)M^3 r +14112M^4 }{2r \left\{48 \left(\sigma_\ell +9\right)r^4  -18 \left(16 \sigma_\ell +221 \right) Mr^3 +\left( 576\sigma_\ell+13465  \right)M^2 r^2 -\left(384\sigma_\ell +19955\right)M^3 r+10962M^4 \right\}} \right].
\end{align}
\end{widetext}
The operator~${\cal L}_H$ is given by Eq.~\eqref{eq:LHGR}. Note that the piece of $\epsilon_2$ has the identical metric perturbation to GR~\cite{Cardoso:2018ptl}, and thus, gives no contributions in the source term. The odd-parity gravitational perturbation of the $i$th order in the low-frequency expansion is governed by
\begin{align}
    {\cal L}_h\left[h^{(i)}\right]=\epsilon_1 {\cal S}_h^{\epsilon_1}h^{(i)}+\epsilon_2 {\cal S}_h^{\epsilon_2}h^{(i)},\label{eqforhinEFT} 
\end{align}
with 
\begin{widetext}
\begin{align}
    &{\cal S}_h^{\epsilon_1}:=\frac{2304M^8 \left(11M-8r\right) }{r^{10}} \left[\frac{d}{dr}+\frac{144\sigma_\ell r^3-18\left(31 \sigma_\ell+24\right)M r^2+\left(545 \sigma_\ell+1646\right)M^2r-1584M^3}{18\left(11M-8r\right)r^3 f^2}\right],\\
    &{\cal S}_h^{\epsilon_2}:=-\frac{36864 \sigma_\ell M^8 r }{r^{10}} \left[\frac{d}{dr}-\frac{\left(\sigma_\ell+14\right)r-32M}{8r^2f}\right].\nonumber
\end{align}
\end{widetext}
Here, ${\cal L}_h$ is given by Eq.~\eqref{eq:LhGR}.

\subsubsection{General solution within perturbative expansions}
We expand $H^{(i)}$ and $h^{(i)}$ in $\epsilon_k$ to ${\cal O}(\epsilon_k)$:
\begin{align}
    H^{(i)}\simeq&  H^{(0,i)}+\epsilon_1 H^{(1,i)},\\
     h^{(i)}\simeq&  h^{(0,i)}+\epsilon_1 h_{\epsilon_1}^{(1,i)}+\epsilon_2 h_{\epsilon_2}^{(1,i)}.
\end{align}
Equations~\eqref{eqforHinEFT} and~\eqref{eqforhinEFT}  reduce to Eq.~\eqref{eqforHh} at ${\cal O}(\epsilon_k^0)$, and 
\begin{align}
{\cal L}_H\left[H^{(1,i)}\right]=&{\cal S}_H^{\epsilon_1} H^{(0,i)},\\
{\cal L}_h\left[h_{\epsilon_1}^{(1,i)}\right]=&{\cal S}_h^{\epsilon_1} h^{(0,i)},\\
{\cal L}_h\left[h_{\epsilon_2}^{(1,i)}\right]=&{\cal S}_h^{\epsilon_2}h^{(0,i)},
\end{align}
at ${\cal O}(\epsilon_k)$. 
The general solutions take the form of Eq.~\eqref{eq:HhGR} at ${\cal O}(\epsilon_k^0)$ and Eqs.~\eqref{eq:H1i} and~\eqref{eq:h1i} at ${\cal O}(\epsilon_k)$.

\subsubsection{Horizon-regular solution}
Imposing a regularity condition at the BH horizon~$r=2M(1+5\epsilon_1 /8)$ specifies the linear perturbation with the ingoing Eddington-Finkelstein coordinates~$(v,r,\theta,\varphi)$, where
\begin{align}
    v:=&t+r_{\rm EFT}^*,\\
    r_{\rm EFT}^*:=&r+2M\ln \left(\frac{r}{2M}-1\right)+\epsilon_1 \delta  r^* +{\cal O}\left( \epsilon_1^2 \right).\nonumber
\end{align}
Here, we do not provide the explicit form of $\delta r^*$. Eliminating logarithmically divergent terms in $X_\omega \left(1+i \omega r_{\rm EFT}^*\right)$ around the horizon with $X_\omega=X_\omega(r)$ representing a collection of $H$ and $h$, we obtain horizon-regular solutions, thereby determining the integration constants~$\mathbb{I}_{H/h}^{(i,j)}$ in terms of $\mathbb{E}_{H/h}^{(i,j)}$. To determine the CTRF~\eqref{eq:CTRFs}, we use Eqs.~\eqref{eq:E0HE1H} and~\eqref{eq:E0hE1h}.

\bibliography{apssamp}

\end{document}